\documentclass[amsmath,amssymb,twocolumn]{revtex4}

\usepackage[dvips]{graphicx}

\usepackage{bm}

\begin{document}

\title{Zigzag edge modes in $Z_2$ topological insulator: reentrance and completely flat spectrum}

\author{Ken-Ichiro Imura$^{1,2}$, Ai Yamakage$^1$, Shijun Mao$^{1,3}$, Akira Hotta$^1$ 
and Yoshio Kuramoto$^1$}
\affiliation{$^1$Department of Physics, Tohoku University, Sendai 980-8578, Japan,}
\affiliation{$^2$Department of Quantum Matter, AdSM, Hiroshima University, Higashi-Hiroshima 739-8530, Japan,}
\affiliation{$^3$Department of Physics, Tsinghua University, Beijing 100084, P.R. China.}


\begin{abstract}
The spectrum and wave function of helical edge modes in $Z_2$ topological insulator
are derived on a square lattice using Bernevig-Hughes-Zhang (BHZ) model.  
The BHZ model 
is characterized by a "mass" term 
$M({\bm k}) = \Delta -B{\bm k}^2$.
A topological insulator realizes when the parameters $\Delta$ and $B$ fall on the regime,
either $0 < \Delta /B < 4$ or $4 < \Delta /B < 8$.
At $\Delta /B = 4 $, which separates the cases of positive and negative (quantized) 
spin Hall conductivities, 
the edge modes show a corresponding change that depends on the edge geometry.
In the $(1,0)$-edge, the spectrum of edge mode remains the same against change of $\Delta /B$, 
although the main location of the mode moves from the zone center for $\Delta /B < 4$, 
to the zone boundary for $\Delta /B > 4$ of the 1D Brillouin zone.
In the $(1,1)$-edge geometry,
the group velocity at the zone center changes sign at $\Delta /B = 4$ where
the spectrum becomes independent of the momentum, {\it i.e.} flat, over the whole 1D Brillouin zone.
Furthermore, for $\Delta/B < 1.354$, the edge mode starting from the zone center vanishes 
in an intermediate region of the 1D Brillouin zone, but reenters near the zone boundary, 
where the energy of the edge mode is marginally below the lowest bulk excitations.  
On the other hand, the behavior of reentrant mode
in real space is indistinguishable from an ordinary edge mode.

\end{abstract}

\date{\today}

\maketitle

\section{Introduction}

Insulating states of non-trivial topological order have attracted much attention
both theoretically and experimentally.
A topological insulator has a remarkable property of being metallic on the surface
albeit insulating in the bulk.
Recently much focus is on a specific type of topological insulators,\cite{HgTe_JPSJ}
which are said to be "$Z_2$-nontrivial". \cite{KM_Z2}
The latter occurs as a consequence of interplay between
a specific type of spin-orbit interaction and band structure. \cite{KM_QSH}
Such systems are invariant under time reversal and show Kramers degeneracy.
From the viewpoint of an experimental realization, 
the original idea of Kane and Mele (KM) \cite{KM_QSH,KM_Z2} was often criticized for being unrealistic,
since it relies on a relatively weak spin-orbit coupling in graphene.
In order to overcome such difficulty, 
Bernevig, Hugues and Zhang (BHZ)
proposed an alternative system \cite{BHZ},
which is also $Z_2$-nontrivial but not based on graphene.
BHZ model was intended to describe low-energy electronic properties
of a two-dimensional (2D) layer of HgTe/CdTe quantum well.
Conductance measurement in a ribbon geometry \cite{Laurens} showed that
the system exhibits indeed a metallic surface state, which is also called {\it helical} edge modes.

This paper highlights the spectrum and wave function of such helical edge modes in the BHZ model.
Respecting appropriately the crystal structure of original HgTe/CdTe quantum well, one can safely
implement it as a tight-binding model on a 2D square lattice.\cite{BHZ, HgTe_JPSJ}
Note, however, that in contrast to KM model which can be represented as a purely lattice model,
in BHZ an internal spin-$1/2$ degree of freedom, stemming from the
$s$-type $\Gamma_6$ and $p$-type $\Gamma_8$ orbitals,
resides on each site of the square lattice in addition to the real electronic spin.
We also mention that in the continuum limit with vanishing topological mass term, KM model has 
two valleys ($K$ and $K'$), whereas BHZ has a single valley (at $\Gamma$).
Another idea which we can borrow from graphene study is the sensibility of edge spectrum on
different types of edge structure, i.e., either zigzag of armchair type.
\cite{Neto,W_PhD}
This applies also to the helical edge modes of BHZ topological insulator in a ribbon geometry,
since the edge spectrum is predominantly determined by how the 2D bulk band structure is "projected"
onto the 1D edge axis.
Indeed, the structure of BHZ helical edge modes has been extensively studied in Ref. \cite{HgTe_JPSJ}
in the $(1,0)$-edge geometry and in the tight-binding implementation.
However, in the practical experimental setup, \cite{Laurens} 
this is certainly not the only one which is relevant to determine the transport characteristics 
at the edges.
Here, in this paper our main focus is on the other representative geometry, the $(1,1)$-edge.
In the $(1,1)$-geometry, as a consequence of the specific way how 
"hidden" Dirac cones (or valleys) in the 2D spectrum is projected onto the $(1,1)$-axis, 
edge modes show some unexpected behaviors.

In order to motivate further the present study,
let us first recall the importance of edge modes in 
the quantum Hall state under magnetic field that 
exhibits a finite and quantized (charge) Hall conductivity $\sigma_{xy}^c$.
In realistic samples with a boundary, quantization of Hall conductivity is
attributed to dissipationless transport due to a gapless chiral edge mode.
In contrast to charge Hall effect, a finite {\it spin} Hall current does not need 
breaking of the time reversal symmetry.
In the quantized spin Hall (QSH) effect,
the Chern number in the bulk takes an integral value,
and correspondingly there appears integral pairs of gapless edge modes.
On the other hand,  the
$Z_2$ topological insulator is characterized by an odd number of
gapless modes per edge that are robust against weak perturbations
preserving the time-reversal symmetry.

The existence of such gapless edge states is generally
guaranteed by a general theorem 
under the name of bulk/edge correspondence.
\cite{Wen,HG}
The BHZ model has a convenient feature that the location of the minimum energy gap
can be controlled by changing the parameters in the model.
In particular,
the sign of spin Hall conductivity changes discontinuously as the mass parameter of the model is varied.
Hence, the corresponding change of edge spectrum should provide useful information on
the bulk/edge correspondence.
Furthermore,  understanding of the nature of edge modes under specific edge geometries
should serve as
possible application of topologically protected phenomena in nano-architectonics.
We take the representative cases of 
the $(1,0)$- and $(1,1)$-edges, which we call also the straight and zigzag edges, respectively.

\begin{figure}[htdp]
\includegraphics[width=8 cm]{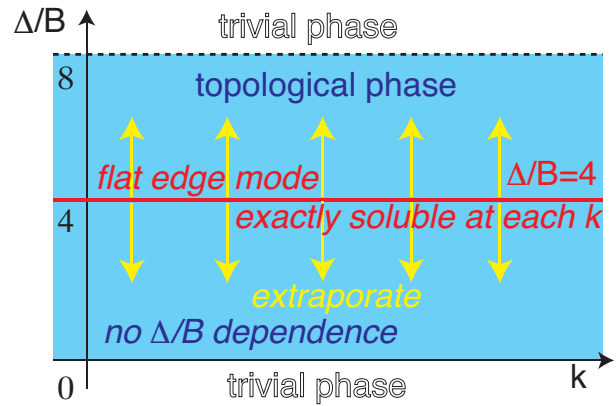}
\caption{Recipe (conceptual) for constructing the exact edge wave function in the zigzag edge
geometry ---
a half-empirical way.}
\label{zig_concept}
\end{figure}

This paper is organized as follows:
In Sec. II, we clarify our motivations to study the lattice version of 
a $Z_2$ topological insulator (BHZ model),
implemented as a nearest-neighbor (NN) tight-binding model.
Explicit form of the BHZ lattice hamiltonian is introduced in Sec. III.
It is demonstrated that
by considering a lattice model, one naturally takes
into account {\it hidden} Dirac cones,
the latter lacking in the analyses based on the continuum Dirac model.
In Secs. IV and V,
we study the detailed structure of gapless edge modes 
under two different edge geometries:
straight and zigzag edges.
Sec. V is the highlight of this paper,
demonstrating that the zigzag edge modes of BHZ lattice model
show unique features.
We first point out that at $\Delta = 4 B$,
a pair of completely flat spectrum appears at $E=0$;
besides the edge wave function can be trivially solved.
We then show, in a half-empirical way, that this analytic exact solution 
at $\Delta = 4 B$ 
can be generalized to the case of an arbitrary $\Delta / B \in [0, 8]$
(this idea is schematically represented in FIG. \ref{zig_concept}).
Using the exact solution thus constructed, we highlight the nature of {\it reentrant} 
edge modes,
another unique feature of the edge modes in the $(1,1)$-edge geometry.
The reentrant edge modes possess two contrasting characters
in real and momentum spaces:
though well distinguished in real space, they live in an extremely small energy
scale in the spectrum.
Sec. VI is devoted to conclusions.
The gapless edge modes of BHZ topological insulator are also
treated in the framework of continuum Dirac model in Appendix A.

\section{Statement of the Problem}

It has been well recognized that the quantized spin Hall conductivity is determined by 
wave functions of Bloch electrons over the entire Brillouin zone. 
On the other hand, only the electronic states near the Fermi level are relevant to 
the change of the Hall conductivity when the Fermi level is shifted.  
Simplified effective models are useful for the latter case since 
various theoretical techniques can be employed in the low-energy range.
In this paper, we work mainly with the lattice version of the BHZ model,
and make some comments in the low-momentum limit.

\subsection{Continuum vs. lattice theories}
Why do we have to go back to a lattice model?
Firstly, because we need to recover the {\it correct} absolute value of spin Hall conductance
$\sigma_{xy}^s$.
The latter is defined as the difference of Hall conductance for up and down 
(pseudo) spins multiplied by $- \hbar /(2e)$:
\begin{equation}
\sigma_{xy}^s = - {\hbar \over 2e}(\sigma_{xy}^\uparrow-\sigma_{xy}^\downarrow).
\end{equation}
In quantized spin Hall (QSH) systems,
the spin Hall conductance is quantized to be an integer in units of $e / (2\pi)$.
\footnote{integer multiple of $2e^2/h$ in the language of charge conductance.}
This is completely in parallel with the quantization of {\it charge} Hall conductance
in units of $e^2/h$ in quantized Hall system.
In both cases, such integers are topological invariants and protected 
against weak perturbations.

On the other hand, if one calculates, using Kubo formula, the contribution
of a single Dirac fermion, 
e.g., of the continuous Dirac model at the $\Gamma$-point \cite{BHZ}
to spin Hall conductance, then
this gives {\it half} of the value expected from the topological quantization.
\cite{Jackiw,Niemi,Redlich,Kenzo}
In order to be consistent,
it is naturally assumed that there must be {\it even} number of Dirac cones.
\cite{HKW}
However, the low-energy effective theory with which we are starting contains
obviously a single Dirac cone.
\cite{BHZ, Haldane}
As we will see explicitly, such trivial discrepancy is naturally resolved 
by considering a lattice version of the BHZ model.

Another aspect motivating us to employ the lattice version of BHZ model is
the fact the idea of an edge states is a real space concept, 
and we need {\it a priori} to go back to real space to give an unambiguous definition to it.
In this paper, we highlight the detailed structure of gapless edge modes under
a specific edge geometry. 
Clearly, introduction of the latter needs a precise description in real space.
Recall also here that edge modes of graphene nano-ribbon exhibit
contrasting behaviors in zigzag and armchair edge geometries:
e.g., the system becomes either metallic or semi-conducting in the armchair geometry,
depending on $N_r \ {\rm mod}\ 3$, with $N_r$ being the number of rows, whereas
a completely flat edge mode appears in the zigzag geometry.
\cite{Neto,W_PhD}
Where does the difference comes from?
In momentum space, the question is how the bulk Dirac cone structure look like 
when viewed from the edge.
Note that in the zigzag geometry the flat edge mode connects
the two Dirac points: $K$ and $K'$, whereas 
in the armchair geometry these two points are projected onto the same point
at the edge.
In a topological insulator, this bulk to edge projection is
even a more subtle issue, since not all the Dirac cones are explicit
(see Table I).
In a sense, zigzag edge in the square lattice BHZ model
(see FIG. \ref{geo_zig})
plays the following double role:
it resembles the zigzag edge in graphene at $\Delta = 4B$,
whereas it may rather resemble the armchair edge at $\Delta = 0$ and 
at $\Delta = 8B$.

\begin{table*}
\caption{Nature of four Dirac cones in the BHZ lattice model.
The four Dirac cones appear at different values of the tuning parameter $\Delta$, 
and at different points of the BZ: $\Gamma$, $X$, $X'$ and $M$.
Away from the gap closing, such Dirac electrons acquires a mass gap.
The sign of such mass gap, together with the chirality $\chi$, 
determines their contribution to $\sigma_{xy}^{(s)}=\pm e^2/h$.
In the table, only their sign is shown. 
The symmetry of the valence orbital is also shown in the parentheses, which is
either, $s$ (inverted gap) or $p$ (normal gap), corresponding, respectively, to the 
parity eigenvalue: $\delta_s=+1$ or $\delta_p=-1$.
The latter is related to $Z_2$ index $\nu$ as
$(-1)^\nu=\prod_{DP}\delta_{DP}$. 
\cite{FuKane}
We also assume $B>0$ with no loss of generality.}
\begin{center}
\begin{tabular}{llcccclcc}
\hline\hline
Dirac points (DP) &\ \ \ & $\Gamma$ & $X_1$ & $X_2$ & $M$ &\ \ \ &
$\sum_{DP}\sigma^{(s)}_{xy} $ & $\prod_{DP}\delta_{DP}$
\\ \hline
$\bm k=(k_x, k_y)$ at the DP && (0, 0) & $(0, \pi /a)$ & $(\pi /a, 0)$ & $(\pi /a, \pi /a)$
\\ \hline
mass gap && $\Delta$ & $\Delta - 4B$ & $\Delta - 4B$ & $\Delta - 8B$ 
\\ \hline
chirality $\chi$ && + & $-$ & $-$ & + &
\\ \hline\hline
$\Delta<0$ && 
$-$ ($p$) & $+$ ($p$) & $+$ ($p$) & $-$ ($p)$ && 
$0$ & +1
\\
$0<\Delta<4B$ && 
$+$ ($s$) & $+$ ($p$) & $+$ ($p$) & $-$ ($p$) && 
$2e^2/h$ & $-1$
\\
$4B<\Delta<8B$ && 
$+$ ($s$) & $-$ ($s$) & $-$ ($s$) & $-$ ($p$) && 
$-2e^2/h$ & $-1$
\\
$8B<\Delta$ && 
$+$ ($s$) & $-$ ($s$) & $-$ ($s$) & $+$ ($s$) && 
0 & +1
\\ \hline\hline
\end{tabular}
\end{center}
\label{4Dpts}
\end{table*}

\subsection{Continuous Dirac model and its boundary conditions}

Spin Hall conductance is a topological quantity determined by the {\it global}
structure of entire 2D Brillouin zone.
The helical edge modes, encoding the same information, 
is, therefore, {\it a priori} not derived
from a local description in the 2D Brillouin zone.
One exception to such a general idea is the study of Ref. \cite{ShenNiu}
(see also Appendix),
in which gapless edge modes are derived from the continuous model in a strip
geometry by simply applying the condition that all the pseudo spin components of 
wave function vanish at the boundary.
\footnote{In the case of graphene (and also KM) zigzag edges, 
we adopt a different boundary condition:
only the $A$ ($B$)-sublattice component of the wave function
vanishes at one (the other) boundary.
For details, see, Ref. \cite{Neto,W_PhD}.}
This implies that information about the helical edge modes is actually
encoded in the original (single) Dirac cone.
This seems to be rather surprising, if one recalls that
in the case of KM model,
the distinction between trivial and non-trivial phases
is made by a relative sign of the mass gap at $K$- and $K'$- points, 
which are, of course, macroscopically separated 
in momentum space.
Here, in the BHZ model, the same distinction is made by relative sign
between the mass ($k^0$-) term and the $k^2$- term added to the Dirac Hamiltonian.

Motivated by this observation,
we investigate the structure of helical edge modes
under different boundary conditions for the periodic BHZ model,
implemented as a square lattice and nearest-neighbor (NN)
tight-binding model model.
In parallel with the arm chair and zigzag edges for graphene,
we consider (a) usually considered $(1,0)$- (straight) boundary \cite{BHZ},
as well as (b) $(1,1)$- (zigzag) boundary for the tight-binding BHZ model.

\subsection{Explicit vs. hidden Dirac cones}

The idea of focusing on Dirac fermions in the description of quantized Hall effect
has appeared in the context of transitions between different plateaus.
For describing the transitions, half-integer quantization is not a drawback,
since the difference of Hall conductance before and after the gap closing is
quantized to be an integer in units of $e^2/h$: $1/2-(-1/2)=1$ or vice versa.
A discrete jump in the Hall conductance across the transition is indeed
consistent with counting based on the emergence of massless Dirac fermions
at the transition \cite{Oshikawa, SM}.

The absolute value of Hall conductance is, on the other hand,
a winding number, and determined by the vortices \cite{TKNN}.
Here, each vortex gives, in contrast to a Dirac cone, an integral contribution to
the Hall conductance (in units of $e^2/h$).
An interesting question is
whether the total Hall conductance, often expressed as a topological invariant
\cite{TKNN},
can be also written as a sum of contributions from emergent Dirac electrons
in the spectrum.
Our empirical answer is yes,\cite{HKW} 
indicating that the number of Dirac electrons
emergent in the spectrum is always even,
reminiscent of
the no-go theorem of Nielsen and Ninomiya in 3+1 dimensions
\cite{nogo}.
It should be noted that here not only explicit Dirac electrons
(gapless for a given set of parameters)
but also hidden Dirac electrons (massive for that set of parameters) 
must be taken into account.
Such massive Dirac electrons are called "spectators" in Ref. \cite{HKW},
in the sense that they are inactive for the transition.
Spectators are indispensable to ensure the correct integer quantization of the
Hall conductance.

\begin{figure}[htdp]
\includegraphics[width=8cm]{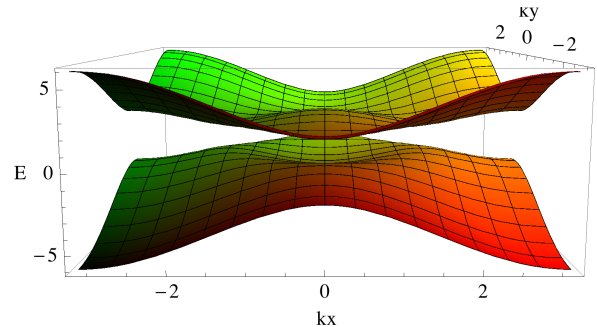}
\caption{Bulk energy spectrum: $E=E(k_x, k_y)$ (vertical axis) of a BHZ lattice model,
here, implemented as a NN tight-binding model on a square lattice:
cf. Eqs. (\ref{dk_NN}).
The spectrum is 
shown over the entire Brillouin zone: $-\pi /a < k_x, k_y < \pi /a$ (horizontal plane).
Parameters are chosen such that $A=B=1$ and $\Delta=2$, i.e., the system is
in the topologically non-trivial phase: $0<\Delta /B <8$.
The spectrum shows a typical wine-bottle structure around the $\Gamma$-point.}
\label{spec_BHZ}
\end{figure}

\section{BHZ models}

\subsection{BHZ model in the long-wave-length limit}

Let us first consider the BHZ model in the long-wave-length limit.
The low-energy effective Hamiltonian,
describing the vicinity of gap closing at $\Gamma =(0,0)$,
is the minimal model to capture the physics of 
a $Z_2$-topological insulator.
This effective Hamiltonian is also contrasting
to the prototypical KM model,
in that the former
describes only a single Dirac cone.
The distinction between the $Z_2$- trivial and non-trivial phases
is, therefore, made by adding a $k^2$-term to the usual Dirac Hamiltonian.
The explicit form of BHZ Hamiltonian is implemented by the following $4\times 4$
matrix:
\begin{eqnarray}
H (\bm k)&=&\left[
\begin{array}{cc}
h(\bm k)  &   0 \\
0  &   h^*(-\bm k) \\
\end{array}
\right],
\label{Htot}
\end{eqnarray}
where $\bm k=(k_x,k_y)$ is a 2D crystal momentum, here 
measured from the $\Gamma$-point.
The lower-right block $h^*(- \bm k)$ is deduced from $h(\bm k)$
by imposing time reversal symmetry.

The bulk energy spectrum: $E=E(\bm k)$ is then
given by solving the eigenvalue equation for the upper-left block $h(\bm k)$
of the $4\times 4$ BHZ Hamiltonian, i.e.,
\begin{equation}
h(\bm k) \psi (\bm k)=E \psi(\bm k).
\label{Psi_k}
\end{equation}
In order to represent $h(\bm k)$ in a compact form,
we introduce a $\bm d$-vector,
$\bm d = ( d_x, d_y, d_z)$,
each component of which is
either an even or an odd function of $\bm k$:
$d_{x,y,z}=d_{x,y,z} (\bm k)$,
whose parity is determined by symmetry considerations.
\cite{BHZ}
As far as the low-energy universal properties in the vicinity of
$\Gamma$-point is concerned, 
we need to keep only the lowest order terms of $d_{x,y,z} (\bm k)$,
and in this long-wave-length limit, they read explicitly,
\begin{eqnarray}
d_x (\bm k) &=& A k_x,\ \ \
d_y (\bm k) = A k_y \\ \nonumber
d_z (\bm k) &=& \Delta-B(k_x^2+k_y^2).
\label{dk}
\end{eqnarray}
Other parameters which appear in Ref. \cite{BHZ}, i.e.,
$C$ and $D$ are set to be zero,
which, however, does not lose any generality.
The bulk energy spectrum is thus determined by diagonalizing
the following "spin Hamiltonian",
$h (\bm k) =\bm d (\bm k) \cdot \bm \sigma$,
where
$\bm \sigma=(\sigma_x, \sigma_y, \sigma_z)$,
are Pauli matrices.
Using the standard representation for $\bm \sigma$, 
$h (\bm k)$ reads explicitly as,
\begin{equation}
h (\bm k) =\left[
\begin{array}{cc}
d_z  &  d_x -i d_y  \\
d_x  +i d_y &  - d_z
\end{array}
\right].
\label{hk}
\end{equation}
Each row and column of Eq. (\ref{hk})
represent an "orbital spin" associated with
the $s$-type $\Gamma_6$ and the $p$-type $\Gamma_8$ orbitals
of the original 3D band structure of HgTe and CdTe.
\cite{HgTe_band}
Then, by choosing the "spin quantization axis" in the direction of
$\bm d$-vector, one can immediately diagonalize the Hamiltonian
$h(\bm k)$, i.e.,
\begin{equation}
h(\bm k) |\bm d (\bm k) \pm\rangle =
\pm d (\bm k) |\bm d (\bm k) \pm\rangle,
\label{dk_pm}
\end{equation}
where the eigenvalue $E(\bm k)=\pm d (\bm k)$ given by,
\begin{equation}
d(\bm k)=|\bm d (\bm k)|=\sqrt{d_x^2+d_y^2+d_z^2}.
\end{equation}
This implies,
\begin{equation}
E(\bm k)^2=\Delta^2+\left( A^2-2B\Delta \right) |\bm k|^2 +B^2 |\bm k|^4,
\label{wine}
\end{equation}
where 
$|\bm k|^2=k_x^2+k_y^2$.
When
$\Delta > A^2/2B \equiv \Delta_0$,
the dispersion relation (\ref{wine}) represents
a wine-bottle structure (FIG. 2),
i.e., $E(|\bm k|)$ shows a minimum at a finite value of $|\bm k|$.
At the critical value $\Delta_0=A^2/2B$,
the density of states shows van Hove singularity.

\begin{figure}[htdp]
\includegraphics[width=8 cm]{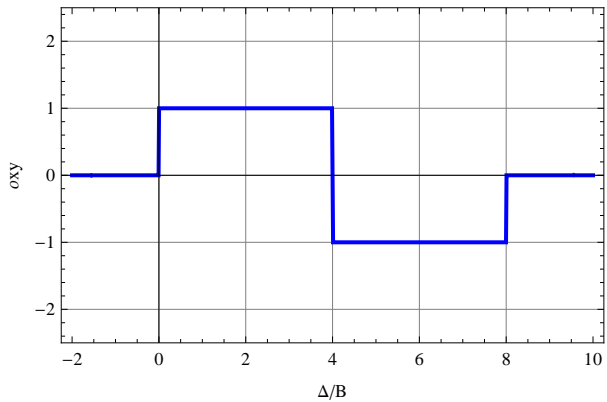}
\caption{$\sigma_{xy}^{(s)}$ in units of $e/2\pi$,
obtained by numerically evaluating the $\bm k$-integral in Eq. (\ref{TKNN}),
plotted as a function of $\Delta /B$.}
\label{sigma_xy}
\end{figure}

\subsection{BHZ model on square lattice and Dirac-cone
interpretation}

Lattice version of the BHZ model is implemented as a
tight-binding Hamiltonian.
To construct such a Hamiltonian explicitly,
we replace linear and quadratic dependences in $h(\bm k)$ 
on $k_x$ and $k_y$ as in Eqs. (\ref{dk}),
by a function 
which has the right periodicity of the square lattice.
This can be implemented as,
\begin{eqnarray}
& d_x (\bm k) \rightarrow {A\over a} \sin (k_x a),
\nonumber \\
& d_y (\bm k) \rightarrow {A\over a} \sin (k_y a),
\nonumber \\
& d_z (\bm k) \rightarrow \Delta - {2B\over a^2}\left[2-\cos (k_x a) - \cos (k_y a)\right],
\label{dk_NN}
\end{eqnarray}
where $a$ is the lattice constant.
Eqs. (\ref{dk_NN}) corresponds to regularizing the effective Dirac
model on a square lattice with only nearest-neighbor (NN) hopping.
In this setup, i.e.,  Eqs. (\ref{Htot}), (\ref{hk}) together with Eqs. (\ref{dk_NN}), 
the lattice version of BHZ model acquires four gap closing
points shown in TABLE I, if one allows the original mass parameter $\Delta$ 
to vary beyond the vicinity of $\Delta=0$.
The new gap closing occurs at different points in the Brillouin zone
from the original Dirac cone ($\Gamma$-point),
namely at
$X_1=(\pi/a,0)$, $X_2=(0,\pi/a)$ and $M=(\pi/a,\pi/a)$.
The gap closing at $M$ occurs at $\Delta=8B$, 
whereas the gap closing at $X_1$ and $X_2$
occurs simultaneously when $\Delta=4B$.

Each time a gap closing occurs, one can re-expand the lattice model
with respect to small deviations of $\bm k$ measured from the gap closing.
The new effective model in the vicinity of such hidden gap closing
falls on the same Dirac form as the original one at the $\Gamma$-point,
up to the $k^2$-term. 
In order to quantify the emergence of such hidden Dirac cones, 
one still needs the following two parameters:
(i) the mass gap $\Delta$ (especially, its sign), and
(ii) the chirality $\chi$.
The latter is associated with the homotopy in the mapping:
$\bm k \rightarrow \bm d (\bm k)$.
Note that in the gap closing at $X_1$ and $X_2$ at $\Delta=4B$,
the role of $k_\pm = k_x \pm i k_y$ is interchanged 
compared with the original Dirac cone at the $\Gamma$-point.
The former (latter) corresponds to $\chi=-1$
($\chi=+1$).
The missing Dirac partner, in the sense of Ref. \cite{nogo}, is found in this way.
Once the explicit form of effective Dirac Hamiltonian in the continuum limit 
is given, one can determine its contribution to $\sigma^{(c,s)}_{xy}$.
In systems with TRS, i.e., of the form (\ref{Htot}),
the contribution from $h(\bm k)$ to 
$\sigma_{xy}^{(c)}=\sigma_{xy}^{\uparrow}+\sigma_{xy}^{\downarrow}$
cancels with that of $h^*(-\bm k)$.
On the other hand, their contribution to
$\sigma_{xy}^{(s)}=-(e/2\pi)(\sigma_{xy}^{\uparrow}-\sigma_{xy}^{\downarrow})$,
remains finite, takes a half-integral value, $\pm 1/2$ in units of
$e/2\pi$ ($2e^2/h$ in conductance).
Contribution to $\sigma^{(s)}_{xy}$ from a Dirac point
with a mass gap $\Delta$ and chirality $\chi$ is,
\begin{equation}
\sigma_{xy}^{(s)}=\chi\ {\rm sign}(\Delta)\ {1\over 2}\ {e\over 2\pi},
\label{1/2}
\end{equation}
provided that the Fermi energy is in the gap.
This can be verified explicitly by applying the Kubo formula to the continuum model.

As mentioned earlier, such counting based on the continuum Dirac model,
is known to describe correctly a discrete jump of $\sigma^{(c)}_{xy}$
in the quantum Hall case.
Here, we apply the same logic to QSH case.
Imagine that one observes the evolution of $\sigma^{(s)}_{xy}$,
starting with the trivial insulator phase, where $\sigma^{(s)}_{xy}=0$,
and varying the mass parameter $\Delta$.
The Fermi energy is always kept in the gap unless there appears a 
Dirac cone. 
Each time such a gap closing occurs, $\sigma^{(s)}_{xy}$ shows a
discrete change, which can be attributed to
the above Dirac fermion argument.
In the present model, one can verify explicitly that this is indeed the case.

If one evaluates the spin Hall conductance from TKNN formula,
\cite{TKNN}
$\sigma_{xy}^{(s)}$ allows for the following representation,
in terms of Berry curvature
integrated over the entire Brillouin zone:
\begin{equation}
\sigma_{xy}^{(s)}=
{e\over 8\pi}\int_{BZ} {d^2 \bm k\over 4\pi}
{\partial \bm d \over \partial k_x}\times
{\partial \bm d \over \partial k_x}\cdot {\bm d \over d^3},
\label{TKNN}
\end{equation}
where $d=|\bm d|$ and $\bm d= \bm d (\bm k)$ is given, e.g., by Eq. (\ref{dk_NN}).
For such an explicit choice of $\bm d (\bm k)$,
Eq. (\ref{TKNN}) is evaluated numerically, and plotted as a function of $\Delta/B$
in Fig. \ref{sigma_xy}. 
When $\bm d (\bm k)$ is given by Eq. (\ref{dk_NN}),
the plotted curve 
(the solid curve shown in blue in Fig. \ref{sigma_xy}, which looks practically like steps)
is comparable with the column $\sum_{DP} \sigma^{(s)}_{xy}$
of TABLE I.
Note that the absolute value of $\sigma_{xy}^{(s)}$ is susceptible of 
the concrete implementation of $\bm d (\bm k)$ over the entire Brillouin zone,
whereas its parity (whether it is even or odd) in units of $e/(2\pi)$ remains the same.
As well known, the latter determines system's $Z_2$-property. \cite{KM_Z2, FuKane}

We have seen that $\sigma_{xy}^{(s)}$
takes a finite value $\pm e/2\pi$
when $0<\Delta /B<8$, i.e., the system is in the topological (inverted gap) phase.
This is also consistent with the gapless edge picture,
in which 
the spin Hall conductance of twice the unit of quantum conductance $e^2/h$
is attributed to two channels of edge modes,
which form a pair of Kramers partners.
The apparent half-integer quantization at the $\Gamma$-point, 
in the sense of Eq. (\ref{1/2}),
is compensated by the contribution from missing Dirac partner(s),
and as a result, is indeed shifted 
by one-half, replaced by an expected integral quantization.

\begin{figure}[htdp]
\includegraphics[width=6 cm]{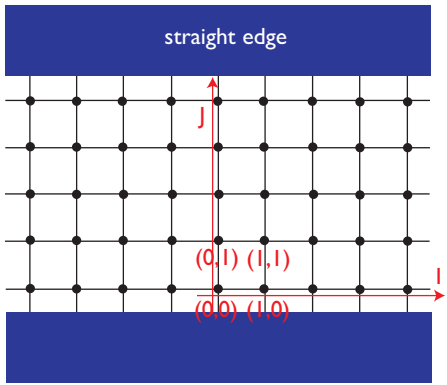}
\caption{Straight edge geometry. The two boundaries of the strip ($=$edges) 
are, here, chosen to be perpendicular to the $(0,1)$-direction. 
In this figure, the number of rows in the strip is $N_r =5$.}
\label{geo_str}
\end{figure}

\section{Straight edge geometry}

Let us first review the behavior of gapless edge modes in the straight edge geometry,
the latter commensurate with the square lattice,
and can be chosen either normal to the $(1,0)$- or to the $(0,1)$-direction (as in FIG. \ref{geo_str}).
Introducing an edge leads to breaking of the translational invariance 
in the direction perpendicular to the edge,
inducing a coupling between Dirac cones.

\begin{figure*}
\begin{minipage}{5.5cm}
\includegraphics[width=5.5cm]{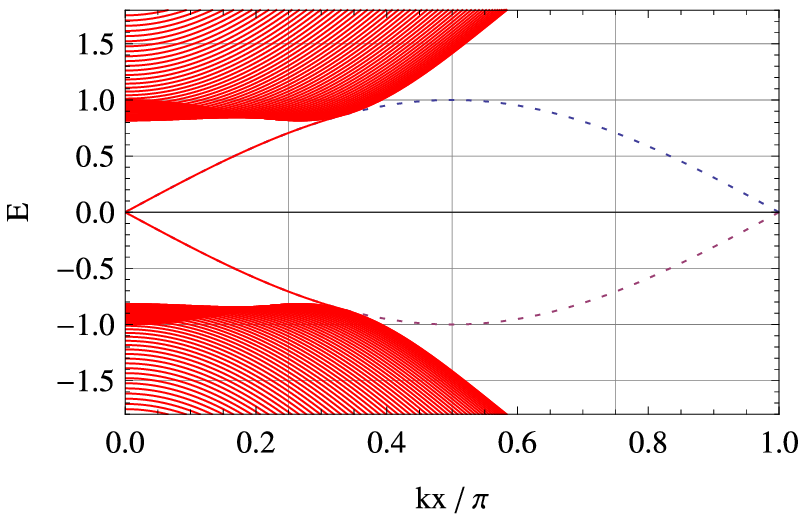}
\end{minipage}
\begin{minipage}{5.5cm}
\includegraphics[width=5.5cm]{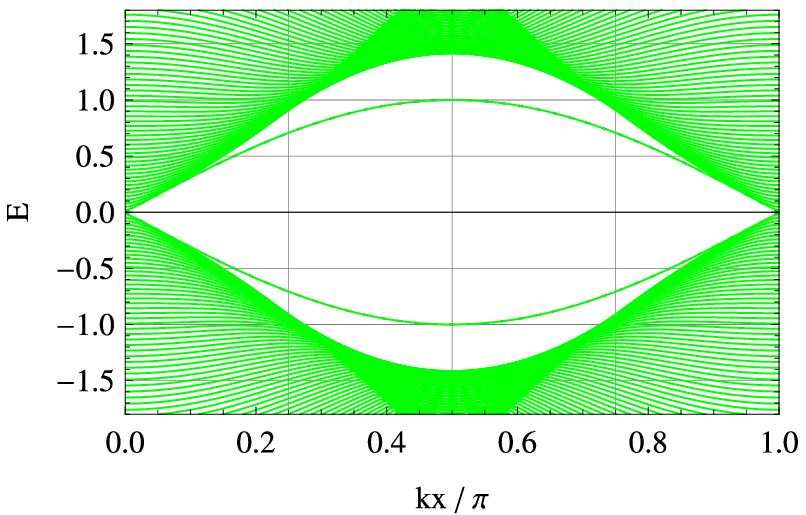}
\end{minipage}
\begin{minipage}{5.5cm}
\includegraphics[width=5.5cm]{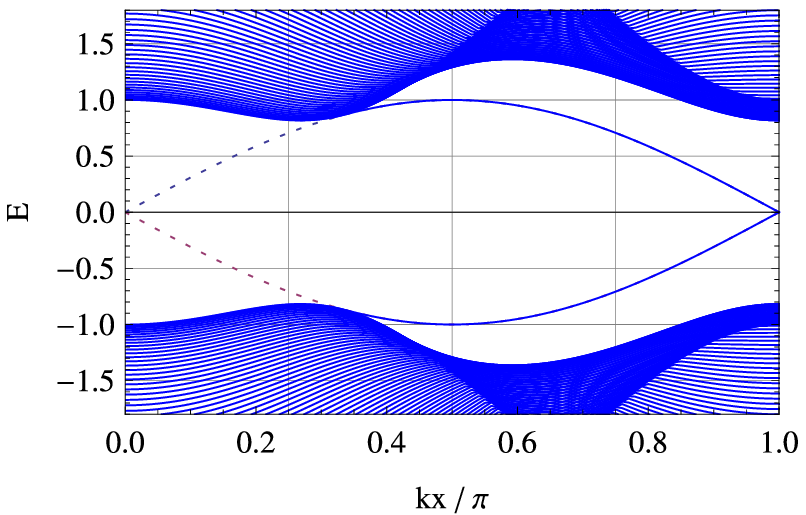}
\end{minipage}

\caption{Energy spectrum (numerical) in the straight edge geometry for
different values of $\Delta$ ($A=B=1$).
The number of rows $N_r$ is, here, chosen to be $N_r =100$.
The dotted curve is a reference, showing the exact edge spectrum
given in Eq. (\ref{spec_th_str}).
Starting with the left panel
$\Delta =  B$ (spectrum shown in red), 
$\Delta = 4 B$ (center panel, spectrum in green), and
$\Delta = 5B$ (right, blue).}
\label{spec_str}
\end{figure*}

\subsection{Effective one-dimensional model}

In the  straight edge geometry shown in FIG. \ref{geo_str}), 
electrons are confined inside a strip between the rows at $y=a$ and $y=N_{r} a$.
The translational invariance along the $x$-axis is still maintained, 
allowing for constructing a 1D Bloch state with a crystal momentum
$k_x$:
\begin{equation}
|k_x,J\rangle=\sum_I e^{i k_x I} |I,J\rangle,
\label{Bloch}
\end{equation}
where
$k_x$ is measured in units of $1/a$
with $a$ being the lattice constant.
$|I,J\rangle=c_{I,J}^\dagger |0\rangle$
is a one-body electronic state localized on site $(I,J)$, and $c_{I,J}^\dagger$
is an operator creating such an electron.
It is also convenient to introduce $c_{k_x,J}^\dagger$, and
express $|k_x,J\rangle$ as
$|k_x,J\rangle=c_{k_x,J}^\dagger |0\rangle$.
Naturally, the two creation operators are related
by Fourier transformation similarly to Eq. (\ref{Bloch}), i.e., 
$c_{k,J}^\dagger=\sum_I e^{i k I} c_{I,J}^\dagger$.

In order to introduce the edges,
it is convenient to rewrite the BHZ tight-binding Hamiltonian
in terms of the hopping between neighboring rows.
Let us first consider
the BHZ tight-binding Hamiltonian in real space:
\begin{eqnarray}
H &=& \sum_{I,J}\left[
(\Delta-4B)\sigma_z c_{I,J}^\dagger c_{I,J}
\right.
\nonumber \\
&+& \left.\left\{
\Gamma_x c_{I+1,J}^\dagger c_{I,J} +
\Gamma_y c_{I,J+1}^\dagger c_{I,J} + h.c.
\right\}
\right],
\label{ham_BHZ}
\end{eqnarray}
where $2\times 2$ hopping matrices, $\Gamma_x$ and $\Gamma_y$,
are given explicitly as,
\begin{equation}
\Gamma_x = - i {A\over 2}\sigma_x + B \sigma_z,\ \ 
\Gamma_y = - i {A\over 2}\sigma_y + B \sigma_z.
\label{gamma_xy}
\end{equation}
In order to rewrite it in terms of the Bloch state, Eq. (\ref{Bloch}), 
or equivalently, in terms of the corresponding
creation and annihiration operators, $c_{k_x,J}^\dagger$ ($c_{k_x,J}$),
we perform Fourier transformation in the ($x$-) direction along the edge.
Eq. (\ref{ham_BHZ}) thus rewrites,
\begin{equation}
H=\sum_{k_x,J} 
\left[
D(k_x) c_{k_x,J}^\dagger c_{k_x,J}+
\left\{ 
\Gamma_{J+1,J} c_{k_x,J+1}^\dagger c_{k_x,J} + h.c.
\right\}
\right]
\label{ham_str}
\end{equation}
where $D (k_x)$'s are diagonal (on-row) components, which read explicitly,
\begin{eqnarray}
D (k_x) &=& A \sigma_x \sin k_x + 
\left\{\Delta - 2B (2 - \cos k_x)\right\} \sigma_z
\nonumber \\
 &\equiv& A \sigma_x \sin k_x +\Omega (k_x) \sigma_z.
\label{dkx}
\end{eqnarray}
$\Gamma_{J+1,J}$ represents a hopping amplitude in the $y$-direction, i.e.,
between neighboring rows.
Inside the strip, i.e., for $J=1,\cdots,N_{r}$,
these amplitudes take the same value as in the bulk, 
given in Eqs. (\ref{gamma_xy}),
i.e., 
\begin{equation}
\Gamma_{J+1,J}=\Gamma_y= - i {A\over 2}\sigma_y + B \sigma_z.
\label{gamma_str}
\end{equation}

In the tight-binding implementation,
a strip geometry can be introduced by switching off all the hopping amplitudes
connecting sites on the edge of the sample to the exterior of the sample.
In our straight edge geometry,
such outermost rows are located at $J=1$ and $J=N_{r}$.
We turn off all the hopping amplitudes from $J=1$ to $J=0$,
and the ones from $J=N_r$ to $J=N_r +1$, i.e.,
\begin{equation}
\Gamma_{1,0}=\Gamma_{N_{r}+1,N_{r}}=0.
\label{bc_tb}
\end{equation}
This boundary condition,
(i) breaks the translational invariance in the $y$-direction,
and
(ii) restricts the Hamiltonian matrix into $N_{r}\times N_{r}$ blocks.

\begin{figure}[htdp]
\includegraphics[width=8 cm]{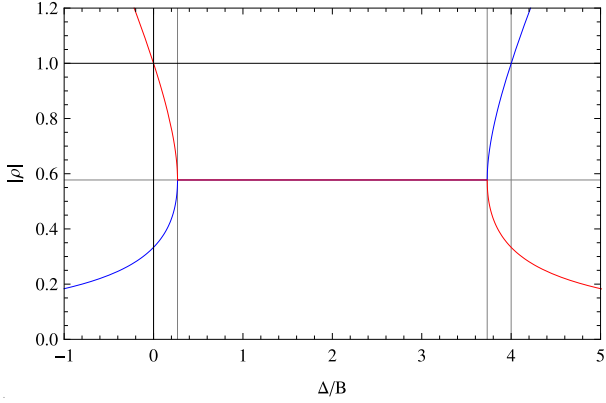}
\caption{$|\rho_1|$ and $|\rho_2|$ plotted as a function $\Delta /B$
in the limit $k_x \rightarrow 0$.}
\label{sol_str1}
\end{figure}

\subsection{Spectrum and wave function}

Let us construct the eigenvector of the straight edge Hamiltonian,
Eq. (\ref{ham_str}), with an eigenenergy $E$.
Since Eq. (\ref{ham_str}) is already diagonal w.r.t. $k_x$,
we diagonalize Eq. (\ref{ham_str}) for a given $k_x$, to find 
the energy spectrum $E=E(k_x)$.
The corresponding eigenvector is thus specified by $E$ and $k_x$,
and takes generally the following form:
\begin{equation}
|E, k_x\rangle=\sum_j \psi_j (E,k_x) |k_x,j\rangle,
\end{equation}
where $\psi_j (E,k_x)$ is a $2\times 2$ spinor specifying the amplitude 
and the pseudo spin state of eigenvector on row $j$.
One might rather regard,
\begin{equation}
\Psi=
\left[\begin{array}{c}
\psi_1\\ \psi_2\\ \psi_3\\ \vdots
\end{array}
\right],
\label{wf}
\end{equation}
as the wave function of the corresponding eigenstate.
The eigenvalue equation, 
\begin{equation}
H |E, k_x\rangle= E (k_x) |E, k_x\rangle,
\end{equation}
can be rewritten, in terms of the $\psi_J (E,k_x)$'s, 
in the form of a recursive equation:
\begin{equation}
D(k_x) \psi_j + \Gamma_y \psi_{j+1} + \Gamma_y^\dagger \psi_{j-1} = E \psi_j.
\label{rec_str}
\end{equation}

All the information on
the spectrum and the wave function of both the extended bulk states and 
the localized edge states is encoded in Eq. (\ref{rec_str})
and the boundary condition which we will specify later.
Since the recursive relation, Eq. (\ref{rec_str}), is linear,
its eigenmodes take the form of a geometric series:
\begin{equation}
\psi_j=\rho^j \psi_0,
\label{ser_rho}
\end{equation}
where
$\rho$ is a solution of the characteristic equation
which we will derive later.
If $|\rho|<1$, Eq. (\ref{ser_rho}) may represent an edge mode.
Since the recursive relation, Eq. (\ref{rec_str}), is of second order,
its characteristic equation becomes a quadratic equation,
giving two solutions for $\rho$, say, $\rho=\rho_{1,2}$.
On the other hand, our recursive equation has also a $2\times 2$ matrix form,
we first have to solve a (reduced) eigenvalue equation for $\Psi_0$,
assuming that $\rho$ is given.
The reduced eigen value equation for $\Psi_0$ reads,
\begin{equation}
\left[
D(k_x) + \rho \Gamma_y + {1\over \rho} \Gamma_y^\dagger
\right] \psi_0 = E \psi_0.
\end{equation}
Using Eqs. (\ref{gamma_xy}) this can be also rewritten as,
\footnote{
The bulk solutions of Eq. (\ref{rec_str}) corresponds to the choice,
$\rho=e^{i k_y}$, or $|\rho|=1$ 
which is consistent with the Bloch theorem.
In the strip geometry with the periodic boundary condition, 
$k_y$ takes discrete values.
With the open boundary condition relevant to the actual strip, 
$k_y$ is no longer a good quantum number.
However, if the width of the strip is large,
one may roughly interpret the 1D energy spectra in the strip geometry as 
composed of the many slices of bulk energy spectrum
at different values of $k_y$.
In addition,  a pair of edge modes appear as a characteristic feature of the nontrivial topological property.}
\begin{equation}
\left[
D(k_x) + i {A \over 2}
\left({1 \over \rho}-\rho \right) \sigma_y 
+ B\left({1 \over \rho}+\rho \right) \sigma_z
\right] \psi_0 = E \psi_0,
\label{eeq_red}
\end{equation}
where $D(k_x)$ is given in Eq. (\ref{dkx}).
This is a $2\times 2$ eigenvalue equation, and there are generally
two solutions for $E$ and two corresponding eigenvectors for a given $\rho$.
Recall here that for $0<\Delta<4$, our numerical data (FIG. \ref{spec_str}) show 
that the edge spectrum behaves as $E\rightarrow 0$ in the limit of $k_x \rightarrow 0$.
Hereafter, we will focus only on such edge solutions.
Since in the same limit, $D (k_x) \rightarrow \Omega(0) \sigma_z$,
Eq. (\ref{eeq_red}) reduces to,
\begin{equation}
\left[
\Omega (0) - {A \over 2}
\left({1 \over \rho}-\rho \right) \sigma_x
+ B\left({1 \over \rho}+\rho \right)
\right] \psi_0 =0.
\label{eeq_0}
\end{equation}
Note that we have multiplied both sides of Eq. (\ref{eeq_red}) by $\sigma_z$.
It is clear from this expression that
$\Psi_0$ can be chosen to be an eigenstate of $\sigma_x$,
i.e., $\Psi_0=|x\pm \rangle$, where
\begin{equation}
|x+ \rangle ={1\over \sqrt{2}}
\left[
\begin{array}{r}
1\\ 
1
\end{array}
\right],\ \
|x- \rangle ={1\over \sqrt{2}}
\left[
\begin{array}{r}
1\\
-1
\end{array}
\right].
\label{x+-}
\end{equation}
If one denotes the eigenvalue of $\sigma_x$ by $s$, as
$\sigma_x \Psi_0 = \pm \Psi_0 \equiv s \Psi_0$, then
$s=1$ corresponds to $\Psi_0=|x+ \rangle$, and
$s=-1$ to $\Psi_0=|x- \rangle$.
Namely, $s$ specifies the  eigenspinors given in Eqs. (\ref{x+-}).
Based on these eigenspinors, one can construct
the total wave function.
Of course, one still needs to do determine the allowed values of $\rho$.
For an eigenstate specified by $s$, $\rho$ must satisfy,
\begin{equation}
\Omega (0) - {A \over 2}
\left({1 \over \rho}-\rho \right) s
+ B\left({1 \over \rho}+\rho \right)=0.
\label{quad}
\end{equation}
For $s=1$,
\begin{equation}
\rho={-\Omega (0)\pm \sqrt{\Omega (0)^2+A^2-4B^2}
\over A+2B}
\equiv \rho_{1,2} (0).
\label{rho12_0}
\end{equation}
In Eq. (\ref{quad}),
if $\rho$ is a solution of this quadratic equation for $s=1$,
then $1/\rho$ satisfies the same equation for $s=-1$.
\cite{HgTe_JPSJ}
Thus the general solution becomes a linear combination of the following 
four basic solutions:
\begin{eqnarray}
\psi_j&=&
\left[ c_{+1} \rho_1 (0)^j + c_{+2} \rho_2 (0) ^j \right] |x+ \rangle 
\nonumber \\
&+&
\left[ c_{-1} \rho_1 (0) ^{-j} + c_{-2} \rho_2 (0) ^{-j}  
\right] |x- \rangle,
\label{sol_0}
\end{eqnarray}
Of course, at this point, this is just a solution at only one single $k$-point, $k_x=0$.
However, a solution in the form of Eq. (\ref{sol_0}), with
$\rho_{1,2} (0)$ given in Eq. (\ref{rho12_0}),
can be easily generalized to satisfy Eq. (\ref{rec_str})
for an arbitrary, finite $k_x$,
after a simple replacement of parameters.
As for the eigenmode of the form of Eq. (\ref{ser_rho}),
one has to solve a reduced eigenvalue equation for $\psi_0$,
and determine $\rho$ such that Eq. (\ref{eeq_red}) is satisfied.
However, since $D (k_x)$ has the structure given in Eq. (\ref{dkx}),
the eigenmodes for $\psi_0$ given as Eq. (\ref{x+-})
remain to be valid for an arbitrary $k_x$.
This might become clearer, if one decomposes Eq. (\ref{eeq_red})
into the following set of equations:
\begin{eqnarray}
A \sigma_x \sin k_x \Psi_0 &=& E \Psi_0
\label{eeq_1}
\\
\left[
\Omega (k_x) \sigma_z
+ i {A \over 2} \left({1 \over \rho}-\rho \right) \sigma_y
\right.
&& \nonumber \\
\left.
+ B\left({1 \over \rho}+\rho \right) \sigma_z
\right] \Psi_0 &=& 0,
\label{eeq_2}
\end{eqnarray}
i.e., Eq. (\ref{eeq_red}) is recovered by adding both sides of Eqs. (\ref{eeq_1}) 
and (\ref{eeq_2}).
The first equation gives the energy dispersion, $E=E_s (k_x)$,
if $\psi_0$ is chosen to be an eigenstate of $\sigma_x$, i.e.,
$\sigma_x \psi_0 = \pm \psi_0 \equiv s \psi_0$,
where
\begin{equation}
E_s (k_x)= s A \sin k_x = \pm A \sin k_x.
\label{spec_th_str}
\end{equation}
Note that this is an exact edge spectrum valid over the entire
range of $k_x$, as far as the edge solution is possible
(see FIG. \ref{spec_str}).
On the other hand, Eq. (\ref{eeq_2}), analogous to Eq. (\ref{eeq_0}),
justifies the previous conjecture: $\psi_0=|x\pm \rangle$.
While, in the characteristic equation for $\rho$, one has to
make the simple replacement :
$\Omega (0) \rightarrow \Omega (k_x)$,
i.e., Eqs. (\ref{eeq_2}) and (\ref{eeq_0}) are identical up to this replacement. 
For $s=1$, the solution for $\rho$ reads,
\begin{equation}
\rho={-\Omega (k_x)\pm \sqrt{\Omega (k_x)^2+A^2-4B^2}
\over A+2B}
\equiv \rho_{1,2} (k_x).
\label{rho_12}
\end{equation}
Correspondingly, a general solution for $\psi_j$ can be constructed as,
\begin{eqnarray}
\psi_j &=&
\left[ c_{+1} \rho_1 (k_x)^j + c_{+2} \rho_2 (k_x) ^j \right] |x+ \rangle 
\nonumber \\
&+& \left[ c_{-1} \rho_1 (k_x) ^{-j} + c_{-2} \rho_2 (k_x) ^{-j} \right] |x- \rangle,
\label{sol_gen}
\end{eqnarray}
where the coefficients $c_{\pm 1,2}$ should be chosen to satisfy the boundary
conditions.
Eq. (\ref{sol_gen}) is 
smoothly connected to Eq. (\ref{sol_0}) in the limit: $k_x \rightarrow 0$.

\subsection{Illustration of edge spectrum}
Three panels of
FIG. \ref{spec_str} show the energy spectrum (edge + bulk) for different values of
$\Delta /B$. 
As for the edge part of the spectrum,
only a part of Eq. (\ref{spec_th_str}) is realized.
In order to determine which part of the spectrum in Eq. (\ref{spec_th_str})
is indeed activated, we discuss below
the case of semi-infinite geometry in some detail.

FIG. \ref{spec_str} also demonstrates one of another specific feature of
straight edge mode that the
main location of the mode moves from the zone center 
for $\Delta < 4 B$, to the zone boundary for $\Delta > 4 B$.
Thus, the group velocity intersecting with the Fermi level reverses
its sign, reflecting the sign change of $\sigma^s_{xy}$ in the bulk.
This can be regarded as the concrete expression of
bulk/edge correspondence in the present case.
\cite{Wen,HG}

What kind of a boundary condition should we apply in Eq. (\ref{sol_gen})?
Suppose that here our system is {\it semi-infinite}, for simplicity,
extended from $j=1$ to $j \rightarrow \infty$.
Such a boundary condition can be applied,
by formally requiring that
the wave function (\ref{sol_gen}) vanishes at $j=0$, i.e.,
\begin{equation}
\psi_0=
\left[
\begin{array}{r}
0\\
0
\end{array}
\right].
\end{equation}
This means that the coefficients $c_{\pm 1,2}$ in Eq. (\ref{sol_gen})
should be chosen to satisfy,
\begin{equation}
c_{+1}+c_{+2}=0,\ \ \
c_{-1}+c_{-2}=0.
\label{req_bc}
\end{equation}
This turns out to be rather an important requirement
for determining the range of validity of the solution given in Eq. (\ref{sol_gen}),
since the wave function $\psi_j$ must be normalizable. 
In Eq. (\ref{sol_gen}),
only the eigenmodes of the form of Eq. (\ref{ser_rho})
with $|\rho|<1$ should be kept in the solution
(to be precise, {\it both} $|\rho_1|$ and $|\rho_2|$ must be smaller than 1).
In a strip geometry,
another solution, consisting of both
$|\rho|>1$, 
describes the edge mode localized 
at the opposite end of the system.

\begin{figure}[htdp]
\includegraphics[width=8 cm]{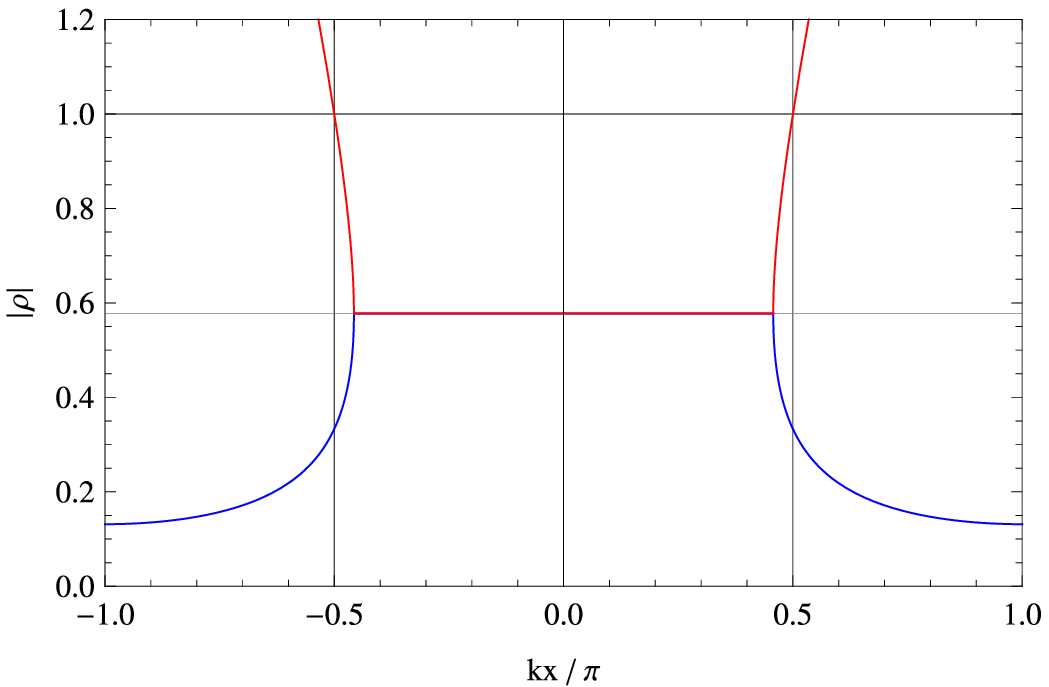}
\includegraphics[width=8 cm]{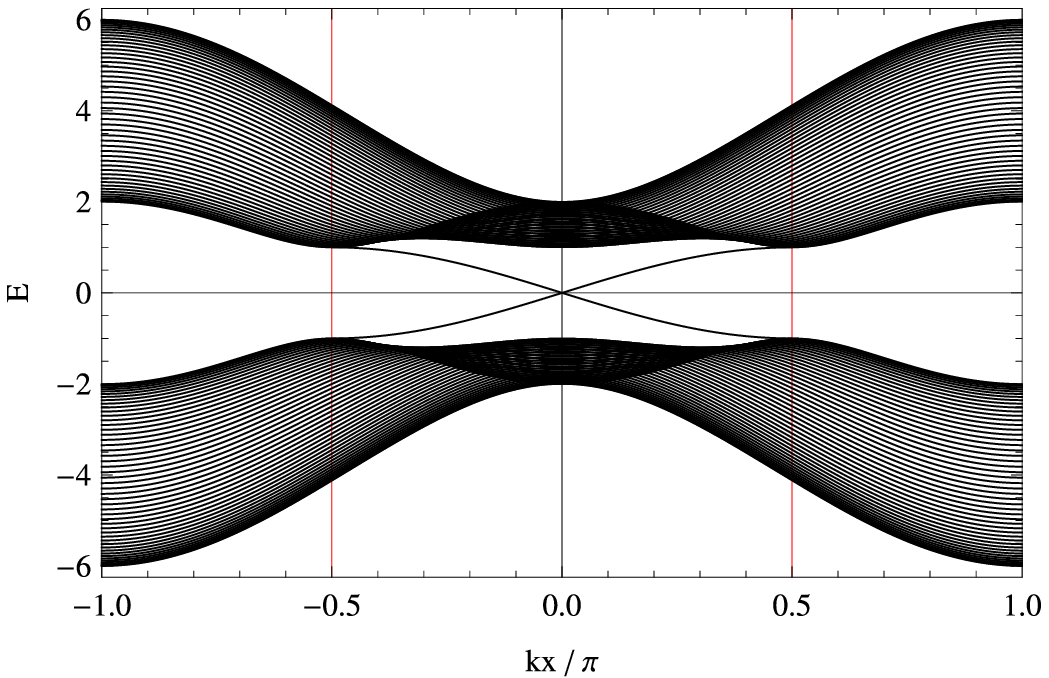}
\includegraphics[width=8 cm]{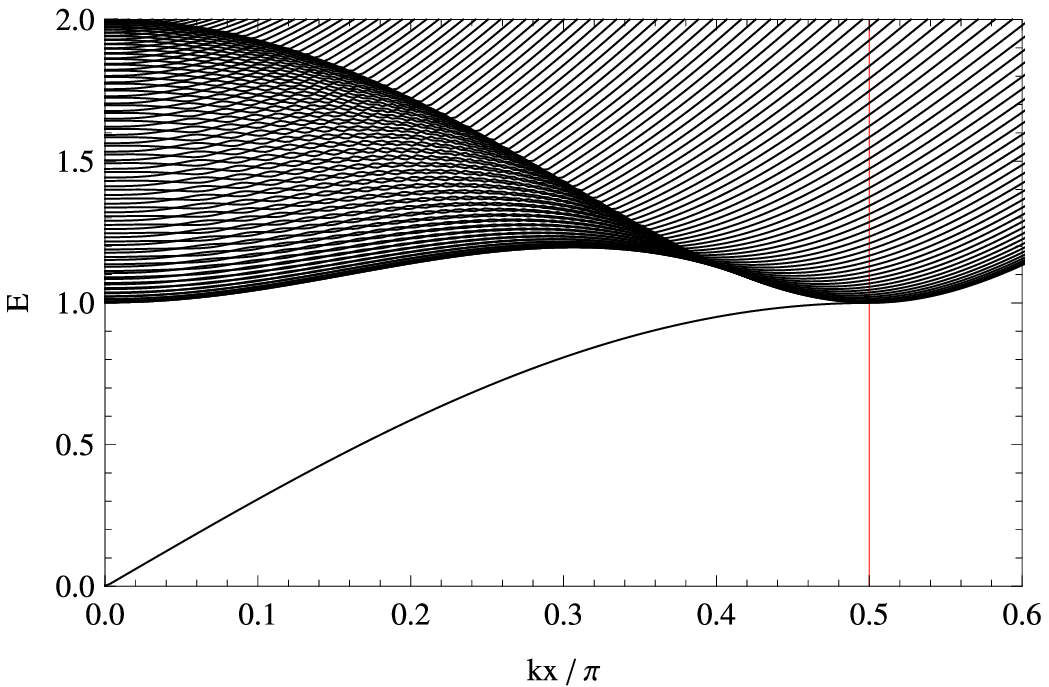}
\caption{Upper panel: $|\rho_1|$ and $|\rho_2|$ plotted as a function of $k_x$
at $\Delta =2$ ($A=B=1$). 
Reference lines are at $k_x/\pi=\pm 0.5$, and at $|\rho|=1$
and $|\rho|=1/\sqrt{3}$ --- as for the latter, cf. Eq. (\ref{rho_flat}).
$|\rho|=1$ corresponds indeed to the point at which 
the edge modes merge with the bulk spectrum (central panel, $N_r =50$).
Lower panel: an enlarged image of the central panel, plotted also 
for a system of larger size: $N_r =100$.}
\label{sol_str2}
\end{figure}

In FIG. \ref{sol_str1}, $|\rho_1|$ and $|\rho_2|$ are plotted 
as a function of $\Delta /B$ in the limit $k_x \rightarrow 0$.
When $0<\Delta /B<4$, both $|\rho_1|$ and $|\rho_2|$ are {\it smaller} than 1,
namely 
both $|1/ \rho_1|$ and $|1/ \rho_2|$ are {\it larger} than 1.
This means that
only the first two terms of Eq. (\ref{sol_gen}),
both corresponding to $|x+\rangle$ ($s=1$),
should be kept in the solution, i.e.,
\begin{equation}
c_{+1} = - c_{+2} \neq 0,\ \ \
c_{-1} = c_{-2} =0.
\label{c+}
\end{equation}
Outside this region, either $|\rho_1|$ or $|\rho_2|$ is larger than 1.
When $|\rho|>1$, 
since this implies automatically $|1/ \rho|<1$,
the eigenmode corresponding to the latter is still compatible with the boundary condition
at $j\rightarrow \infty$.
However, because of the boundary condition at $j=0$, i.e., Eq. (\ref{req_bc}),
when $|\rho_1|>1$ and $|\rho_2|<1$, or vice versa,
the only possible choice for the coefficients $c_{\pm 1,2}$ is
\begin{equation}
c_{+1} = c_{+2} = c_{-1} = c_{-2}=0.
\label{0000}
\end{equation}
Namely, a solution of the type of Eq. (\ref{sol_gen}), or an edge mode crossing
at $k_x=0$ is {\it inexistent}.
This is consistent with the fact that
an edge mode crossing at $k_x=0$ exists only in the region,
$\Delta /B\in [0,4]$ in the straight edge geometry.
(when $\Delta /B \in [4,8]$, the edge modes cross at $k_x=\pi$,
i.e., at the zone boundary, which is also time-reversal symmetric.)

Coming back to the regime in which edge modes are existent, i.e.,
$\Delta /B \in [0,4]$, one can clearly see in FIG. \ref{sol_str1} that there are
two different behaviors --- a flat region where $|\rho_1|$ and $|\rho_2|$
are degenerate, and the remaining part with two branches. 
This is due to the fact that the two solutions for $\rho$ could be either 
both real, or a pair of complex numbers conjugate to each other.
In the latter case, the two solutions have the same absolute value, 
$|\rho_1|=|\rho_2|$, whereas in the present case, 
one can verify that this degenerate value 
is independent of $\Delta$, i.e.,
\begin{equation}
|\rho_1|=|\rho_2|=\sqrt{\left| {A - 2B  \over A + 2B} \right|}.
\label{rho_flat}
\end{equation}
This explains the existence of a 
flat region in FIG. \ref{sol_str1}.
From Eq.(\ref{rho12_0}) we see that the square root becomes pure imaginary for all $k$ provided
$\Delta_- < \Delta < \Delta_+$, where
\footnote{
Real solutions for $\rho$ appear in the regime:
$0 < \Delta < \Delta_-$, and also at the other end.
In the approximation that becomes valid in the small wave number,
this threshold value 
is given by $\Delta_1=A^2/(4B)$.
For $A=B=1$, $\Delta_- = 2-\sqrt{3} = 0.2679\cdots$, whereas,
$\Delta_1$ is, of course, $1/4 = 0.25$.
If one expands Eq. (\ref{delta_pm}) in powers of $A^2/(4 B^2)$,
then at leading order
$\Delta_-$ coincides with $\Delta_1$.}
\begin{equation}
{\Delta_\pm \over B} = 2\left(1\pm \sqrt{1 - {A^2 \over 4 B^2} }\right).
\label{delta_pm}
\end{equation}

At $|\rho|=1$,
the edge solution is expected to merge with the bulk spectrum.
This happens, when
\begin{equation}
\cos\ k_x  = 1- {\Delta \over 2B}.
\label{km}
\end{equation}
Such a behavior becomes clearer by plotting $|\rho|$'s
as a function of $k_x$.
FIG. \ref{sol_str2} illustrates this feature at $\Delta=2$
for $A=B=1$.
One can indeed see that the edge spectra merge with the bulk at
$k_x=k_m$, satisfying Eq. (\ref{km}).
The latter reduces, at this value of $\Delta$, to
$\cos k_m  = 1- \Delta / (2B)=0$, i.e.,
$k_m = \pm \pi / 2$.

It is also instructive to investigate the nature of edge modes
in {\it real} space, i.e., the wave function, and compare it with
the general solution (\ref{sol_gen}).
In numerical experiments, one has to diagonalize 
the $2 N_r\times 2 N_r$ Hamiltonian matrix, equivalent to Eq. (\ref{ham_str}).
An eigen wave function is, therefore, obtained as a $2 N_r$-component 
vector; here, in the straight edge geometry, the latter can be chosen to be real.
The edge wave function is easily identified if it exists, e.g., 
by choosing the lowest-energy eigenmode  $\Psi_0$ 
in the upper band.
By investigating the structure of such an edge wave function, one can explicitly
verify that the eigenmodes are spanned by two eigenspinors given in Eq. (\ref{x+-}).
In repeating such numerical experiments for different $k_x$
and $\Delta /B$,
one can naturally distinguish an edge state from a bulk state
by focusing on the spatial distribution of the wave function.
Here, what deserves much attention is that one can recognize
a one-to-one correspondence
between localizability of the wave function $\Psi_0$
and its spinor structure.

\section{Zigzag edge geometry}

\begin{figure}[htdp]
\includegraphics[width=6 cm]{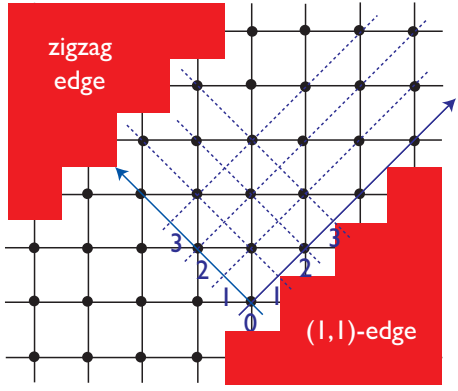}
\caption{Zigzag or $(1,1)$-edge geometry (here, chosen to be normal to the $(1,-1)$-axis),
defined in terms of the original square lattice, on which the new $\tilde{x}$- and 
$\tilde{y}$-axes are superposed in blue (online).
Numbers on these axes are redefined indices $I$ and $J$, i.e.,
$(\tilde{x}, \tilde{y})=(I a/\sqrt{2}, J a/\sqrt{2})$.
The number of rows $N_r$ is here chosen to be, $N_r=6$.
When $N_r$ is even (odd), the two edges are inversion asymmetric (symmetric)
w.r.t. the center of strip.}
\label{geo_zig}
\end{figure}

Let us turn to the case of a different edge geometry,
the zigzag edge geometry, shown schematically in FIG. \ref{geo_zig}.
As mentioned earlier,
the zigzag edge geometry considered here is, in a sense, analogous to 
a more popular edge geometry of graphene ribbon,
named in the same way, but defined on a hexagonal lattice.
Here, on a square lattice, a zigzag edge is introduced, either normal to
$(1,1)$- or $(1,-1)$-direction (as in FIG. \ref{geo_zig}).
Electrons in the zigzag edge geometry are, therefore, confined to
a strip diagonal in the cartesian coordinates, 
say, along the $(1,1)$-direction as in FIG. \ref{geo_zig}.

Intuitively,
say, because the zigzag surface is literally, "rough",
one expects that this edge geometry might have a stronger tendency to trap 
electrons in the vicinity of the boundary.
We show below, on one hand, that 
this intuition from the macroscopic world is still valid
in the microscopic quantum mechanical world.
But just as a result of this stronger tendency to keep the electrons in
its vicinity, on the other hand,
the zigzag edge shows various curious phenomena,
e.g., completely flat edge modes, and the reentrance of edge modes
in $k$-space, etc.

Clearly, the translational invariance along the $(1,1)$-axis 
is maintained, on which $\tilde{x}$-axis is introduced, 
together with the conserved momentum 
$k=k_{\tilde{x}}$ in this direction.
Accordingly,
$\tilde{y}$-axis is chosen to be in the $(1,-1)$-direction.
It may be also useful to {\it redefine} the indices $I$ and $J$ such that
the lattice points in the original square lattice are located at
$(\tilde{x}, \tilde{y})=(I a/\sqrt{2}, J a/\sqrt{2})$, 
where $I$ is an even (odd) integer for $J$: even (odd).
In the zigzag edge geometry, 
the spectrum $E=E (k)$ and the wave function $\Psi$
are determined by
the following recursive equation:
\begin{equation}
(\Delta -4B)\sigma_z \psi_j + \Gamma \psi_{j+1} +\Gamma^\dagger \psi_{j-1} = E \psi_j
\label{rec_zig}
\end{equation}
for $\psi_j$,
analogous to Eq. (\ref{rec_str}) in the straight edge geometry.
In order to derive Eq. (\ref{rec_zig}),
we first rewrote the tight-binding Hamiltonian (15)
in the new labeling, and then considered
a Bloch state along the $\tilde{x}$-axis, analogous to Eq. (\ref{Bloch}),
but with a crystal momentum $k$ conjugate to $\tilde{x}$,
i.e., $k=k_{\tilde{x}}$.
As was the case in Eq. (\ref{rec_str}),
$\Gamma$ describes, here, in Eq. (\ref{rec_zig})
the hopping between adjacent rows, and reads
explicitly as,
\begin{equation}
\Gamma= i{A\over 2} e^{-i k}\sigma_x
- i{A\over 2} e^{i k}\sigma_y
+ 2B\sigma_z\cos k.
\label{gamma_zig}
\end{equation}
Note that here the crystal momentum $k$ is measured in units of $1/(\sqrt{2}a)$
so that the zone boundary is always given by $k=\pi$.
\footnote{Namely, if one compares it to the straight edge case, e.g., in considering
the long-wave-length limit, one has to make the correspondence between
$k\sqrt{2}a$ and $k_x a$.}
To find an edge solution, we first express the solution of Eq. (\ref{rec_zig}),
in the form of a geometric series,
written formally in the same as Eq. (\ref{ser_rho}).
Recall that $\rho$ is (generally) a (complex) number of,
for an edge mode,
amplitude smaller than unity
(with the understanding that the edge mode is localized in the vicinity of $j=0$).
$\psi_0$ is a two component eigenvector of the following reduced eigenvalue equation:
\begin{equation}
\left[
(\Delta -4B)\sigma_z + \rho\Gamma + {1\over\rho}\Gamma^\dagger 
\right]
\psi_{0} = E \psi_0.
\label{reduced}
\end{equation}
From the analogy to the straight edge case, one may express $\Gamma$
as
\begin{equation}
\Gamma= i{A\over 2} c_k (\sigma_x - \sigma_y)
+ {A\over 2} s_k (\sigma_x + \sigma_y)
+2B c_k \sigma_z,
\label{gamma_zig2}
\end{equation}
and rewrite Eq. (\ref{reduced}) into the following explicit form:
\begin{eqnarray}
&&\left[
(\Delta -4B)\sigma_z + 
\left(
\rho + {1\over\rho}
\right)
\left\{
{A\over 2} s_k (\sigma_x + \sigma_y)
+2B c_k \sigma_z
\right\}
\right.
\nonumber \\
&&\left.
+
\left(
\rho - {1\over\rho}
\right)
i{A\over 2} c_k (\sigma_x - \sigma_y)
\right]
\psi_{0} = E \psi_0.
\label{}
\end{eqnarray}
where we $c_k$ and $s_k$ are short-hand notations for, respectively,
$\cos k$ and $\sin k$.
Comparing this form with the straight edge case, one can see that here
one cannot use the same recipe for solving the problem,
i.e., solving the problem at, say, $k=0$ and then extrapolate its solution to general $k$.
It seems not impossible to proceed in that direction and solve the problem analytically,
but here, we choose to take another route, which is much simpler,
to find still an exact solution, but in a half-empirical way.

\begin{figure*}
\begin{minipage}{5.5 cm}
\includegraphics[width=5.5 cm]{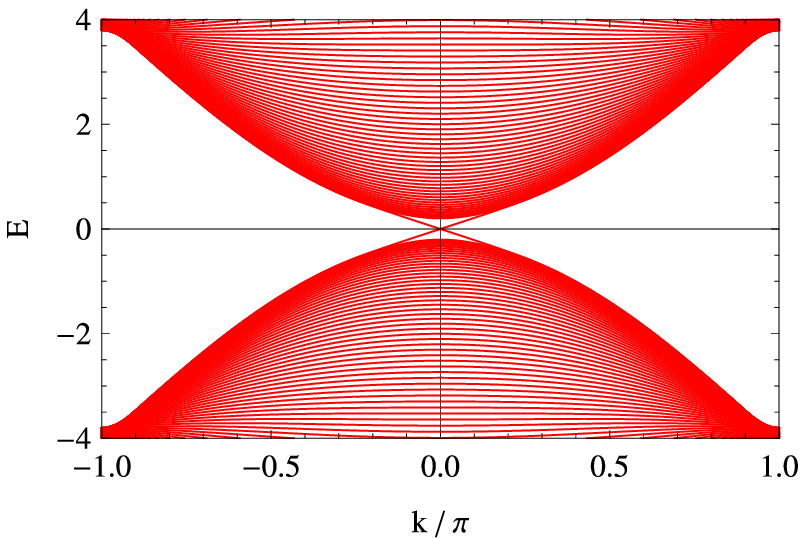}
\end{minipage}
\begin{minipage}{5.5 cm}
\includegraphics[width=5.5 cm]{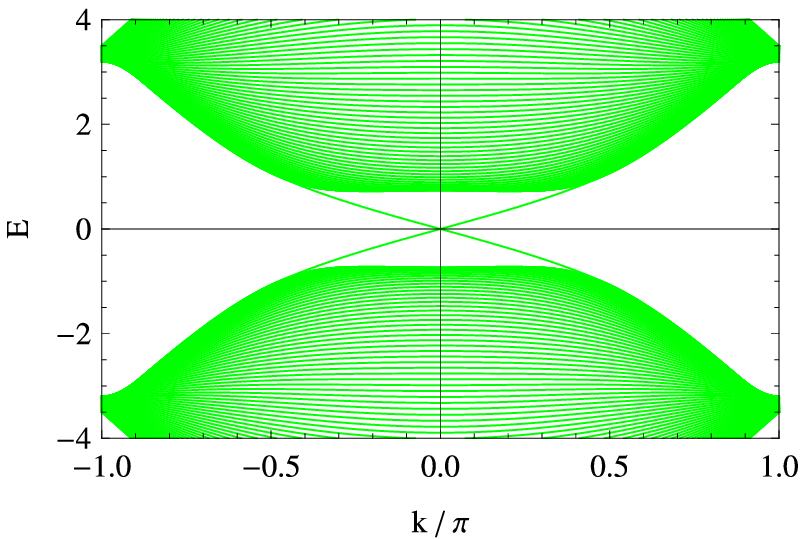}
\end{minipage}
\begin{minipage}{5.5 cm}
\includegraphics[width=5.5 cm]{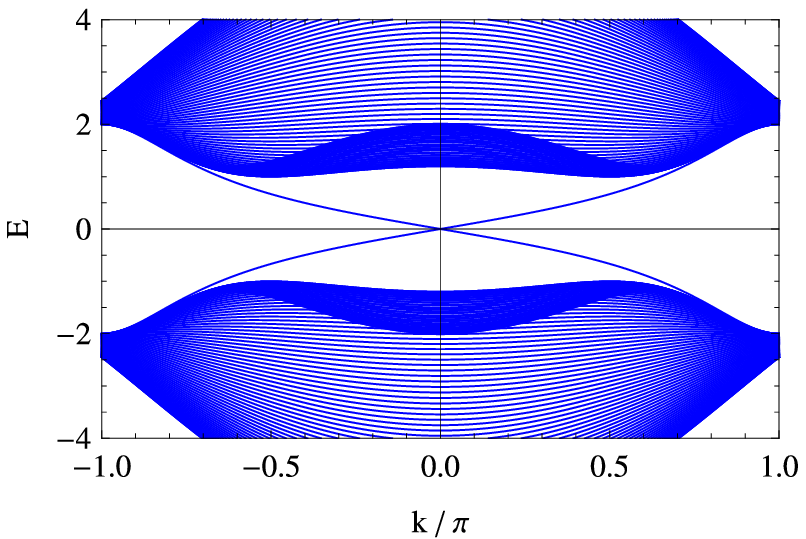}
\end{minipage}
\\
\vspace{0.5 cm}
\begin{minipage}{5.5 cm}
\includegraphics[width=5.5 cm]{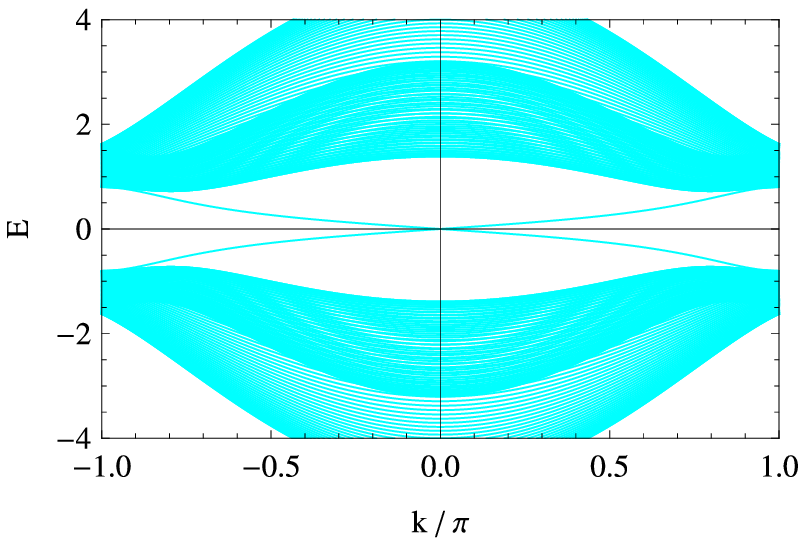}
\end{minipage}
\begin{minipage}{5.5 cm}
\includegraphics[width=5.5 cm]{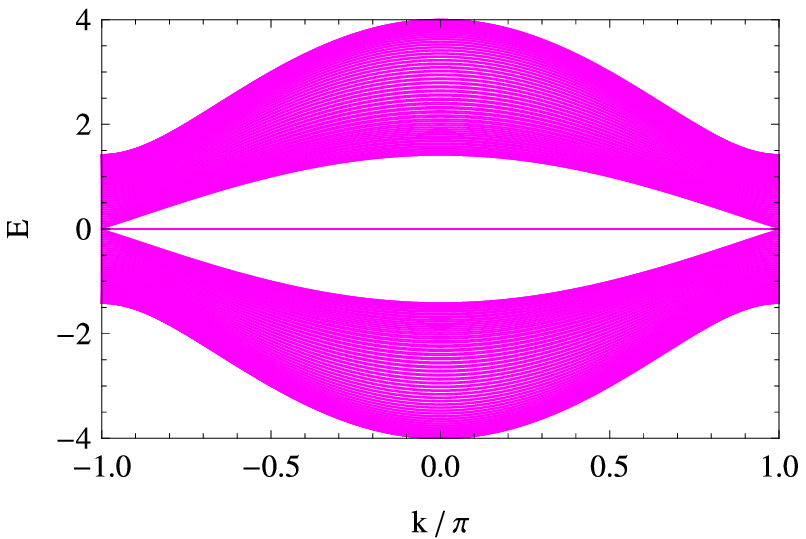}
\end{minipage}
\begin{minipage}{5.5 cm}
\includegraphics[width=5.5 cm]{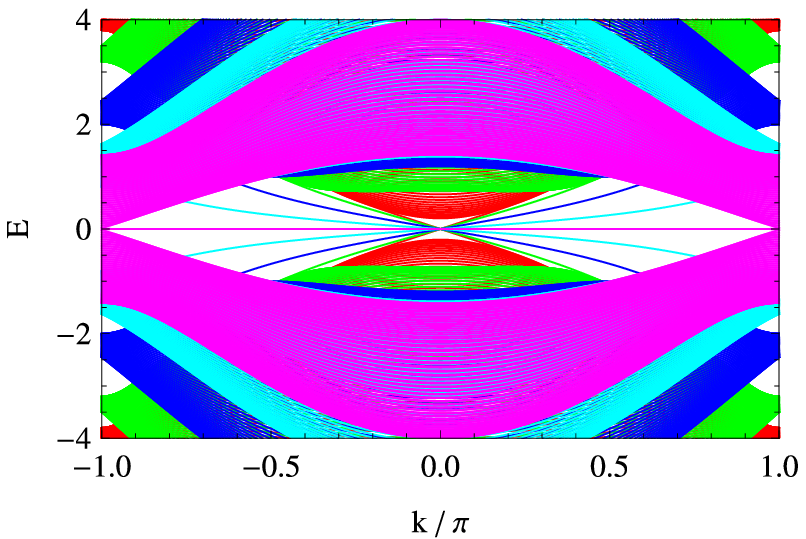}
\end{minipage}

\caption{Energy spectrum in the zigzag edge geometry for
different values of $\Delta$ ($A=B=1$).
Upper-left panel: $\Delta = 0.2 B$ (spectrum shown in red), 
upper-central: $\Delta = 0.8 B$ (spectrum in green),
upper-right: $\Delta = 2 B$ (spectrum in blue),
lower-left: $\Delta = 3.2 B$ (spectrum in cyan), and
lower-central: $\Delta = 4 B$ (spectrum in magenta).
At $\Delta = 4 B$,
the edge modes become completely flat and covers the entire Brillouin zone
(cf. case of graphene in the zigzag edge geometry \cite{W,W_PhD}).
Notice that here the horizontal axe is suppressed to make 
the edge modes legible.
Note that at $\Delta = 4 B$ the bulk spectrum is also gapless;
the completely flat edge modes indeed touch the bulk continuum
at the zone boundary (projection of 2D Dirac cones).
Compare this panel with the corresponding panel of straight edge case:
FIG. (\ref{spec_str}).
The number of rows $N_r$ is here chosen to be $N_r =100$.
These five plots are superposed in the lower-right panel 
to show that the edge spectra at different values of $\Delta$ are,
in contrast to the straight edge case, {\it not} on the same curve.
Even in the long-wave-length limit: $k \rightarrow 0$, their slopes 
are different.}
\label{spec_zig}
\end{figure*}

\subsection{Completely flat edge mode at $\Delta = 4B$}

Some concrete examples of such energy spectrum are shown 
FIG. \ref{spec_zig}.
A pair of gapless edges modes always appear iff $0<\Delta /B <8$.
In contrast to the straight edge case, however,
they appear always in the vicinity of $k=0$, and
intersects at $k=0$.

One of the last panels of FIG. \ref{spec_zig} shows a unique feature
of edge modes in the zigzag edge geometry. 
At $\Delta=4B$, the edge modes become completely flat,
apart from a small finite-size gap around the zone boundary.
We have already seen such flat edge modes in graphene in the case of
zigzag edge geometry (but on a hexagonal lattice).
\cite{W,W_PhD}
In graphene, such flat edge modes connect 1D projection of 
$K$ and $K'$ points via the 1D BZ boundary.
This is a similar behavior to the present case, if one regards the former
as the limit of vanishing intrinsic coupling (or topological mass $\Delta$)
in the KM model.
One of the differences between the two cases is
that here the flat edge modes cover the entire 1D Brillouin zone.

In order to elucidate the nature of flat edge mode at $\Delta = 4B$,
first notice that at this value of $\Delta$ the diagonal terms of 
(the diagonal blocks of the Hamiltonian matrix in) Eq. (\ref{rec_zig}) vanish.
This implies,
as in graphene nano-ribbon in the zigzag edge geometry,
the existence of an eigenstate of the form:
\begin{equation}
\Psi=
\left[
\begin{array}{c}
\psi_1\\
0\\
\psi_3\\
0\\
\psi_5\\
\vdots\\
\end{array}
\right]
\label{psi_2j=0}
\end{equation}
i.e., an eigenvector satisfying $\psi_{2j}=0$ ($j=1,2,\cdots$).
Here, a semi-infinite geometry is implicit
(for approximating a ribbon of sufficiently large width or $N_r$;
our system extended from $j=1$ to $j=N_r$),
with a boundary condition of $\psi_0=0$.
Under this setup, and with the condition of vanishing diagonal matrix elements,
Eq. (\ref{rec_zig}) implies,
\begin{equation}
\Gamma \psi_{2j+1} + \Gamma^\dagger \psi_{2j-1} = 0
\ \ \
(j=1,2,\cdots).
\label{rec_D4B}
\end{equation}
Simultaneously, it should also have a vanishing eigenenergy $E=0$
for consistency.

Clearly, Eq. (\ref{rec_D4B}) has a solution of the form of a geometric series,
here, for $\psi_{2j-1}$ ($j=1, 2, 3, \cdots$):
\begin{equation}
\psi_{2j+1} = \lambda^j \psi_1.
\label{ser_lambda}
\end{equation}
Note that here $\lambda$ plays, roughly, the role of $\rho^2$, but their precise
relation will become clearer when the entire problem is solved.
In order to proceed, we recall that $\Gamma$ can be written explicitly as,
\begin{equation}
\Gamma=
\left[
\begin{array}{cc}
2 B c_k & - {A\over 2}(1-i)(c_k - s_k) \\
{A\over 2}(1+i)(c_k + s_k) & - 2 B c_k
\end{array}
\right].
\label{gamma_zig_mat}
\end{equation}
Then, by assuming a solution of the form of Eq. (\ref{ser_lambda}),
Eq. (\ref{rec_D4B}) can be reduced to the following eigenvalue problem for $\psi_1$:
\begin{equation}
- \Gamma^{-1} \Gamma^\dagger \psi_1 = \lambda \psi_1,
\end{equation}
with the eigenvalues, 
\begin{equation}
\lambda_\pm={A^2 + 8B^2 c_k^2 \pm 2A c_k \sqrt{A^2 s_k^2 + 8B^2 c_k^2}
\over (2 c_k^2 -1) A^2 - 8 B^2 c_k^2}, 
\label{lambda+-}
\end{equation}
and the corresponding eigenvector, $u_\pm$, i.e.,
\begin{equation}
- \Gamma^{-1} \Gamma^\dagger u_\pm = \lambda_\pm u_\pm, 
\end{equation}
given explicitly as,
\begin{equation}
u_\pm =
\left[
\begin{array}{c}
\alpha_\pm (1- i)\\
1
\end{array}
\right].
\label{u+-}
\end{equation}
The coefficient $\alpha_\pm$ is a function of $k$,
which takes precisely the following form:
\begin{equation}
\alpha_\pm (k) = - {1 \over 4B}
\left(
A \tan k \mp \sqrt{A^2 + 8 B^2 \tan^2 k}
\right).
\label{a+-}
\end{equation}
Notice that $\alpha_- = - 1/(2\alpha_+)$, and
the two eigenspinors $u_\pm$ are orthogonal:
$u_-^\dagger u_+ = 0$.
A general solution in the form of Eq. (\ref{psi_2j=0})
can be thus constructed by applying 
$-\Gamma^{-1}\Gamma^\dagger$, recursively, to
\begin{equation}
\psi_1 =c_+ u_+ + c_- u_-,
\end{equation}
and the result is,
\begin{equation}
\psi_{2j+1} =
c_+ \lambda_+^j
\left[
\begin{array}{c}
\alpha_+ (1- i)\\
1
\end{array}
\right]
+
c_- \lambda_-^j
\left[
\begin{array}{c}
\alpha_- (1- i)\\
1
\end{array}
\right].
\label{sol_D4B}
\end{equation}
In this construction, the two eigenvectors $u_\pm$ always have
the form of Eq. (\ref{u+-}).
This feature remains when $\Delta$ is away from $4B$ at which
the edge modes are no longer completely flat, or rather even in the regime in which
the edge spectrum is not flat at all.

Another remark, concerning the behavior of Eq. (\ref{sol_D4B})
is that under the choice of signs in Eq. (\ref{lambda+-}),
$|\lambda_+|$ is always larger than 1, whereas $|\lambda_-|<1$ except at the zone boundary.
This can be easily verified either numerically, i.e.,
by plotting $\lambda_\pm$ as a function of $k$, or
by showing $\lambda-=1/\lambda_+$ using directly the expression 
for $\lambda_\pm$ in Eq. (\ref{lambda+-}).
In numerical simulation for systems of a finite number of rows,
both of these two solutions play a role giving rise to 
a pair of edge solutions.
\footnote{When $N_r$ is odd, one solution, corresponding, say,  to $\lambda_-$,
is localized in the vicinity of $j=1$, and indeed has the form of
Eq. (\ref{psi_2j=0}) with $\psi_{2j+1}$ given
as Eq. (\ref{sol_D4B}) and $c_+=0$.
The other solution, corresponding to $\lambda_+$, has the same structure of
Eq. (\ref{psi_2j=0}) but with $\psi_{2j+1}$
increasing with practically an equal geometric ratio of $\lambda_+$,
and naturally localized in the vicinity of the other edge: $j=N_r$.
On the other hand, when $N_r$ is even, 
the eigenmode of the system becomes a linear combination of the above
two types of solutions.
This even/odd feature occurs only at $\Delta$ precisely equal to $4B$,
since at this value of $\Delta$ where the edge spectrum becomes completely flat,
the two (generally) counter-propagating edge modes
acquire the same (zero) group velocity, and get mixed.}

\begin{figure}[htdp]
\includegraphics[width=8 cm]{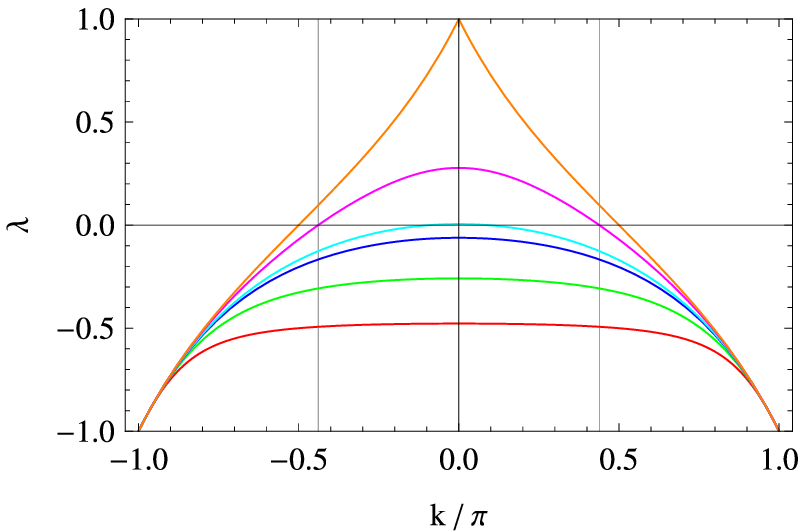}
\includegraphics[width=8 cm]{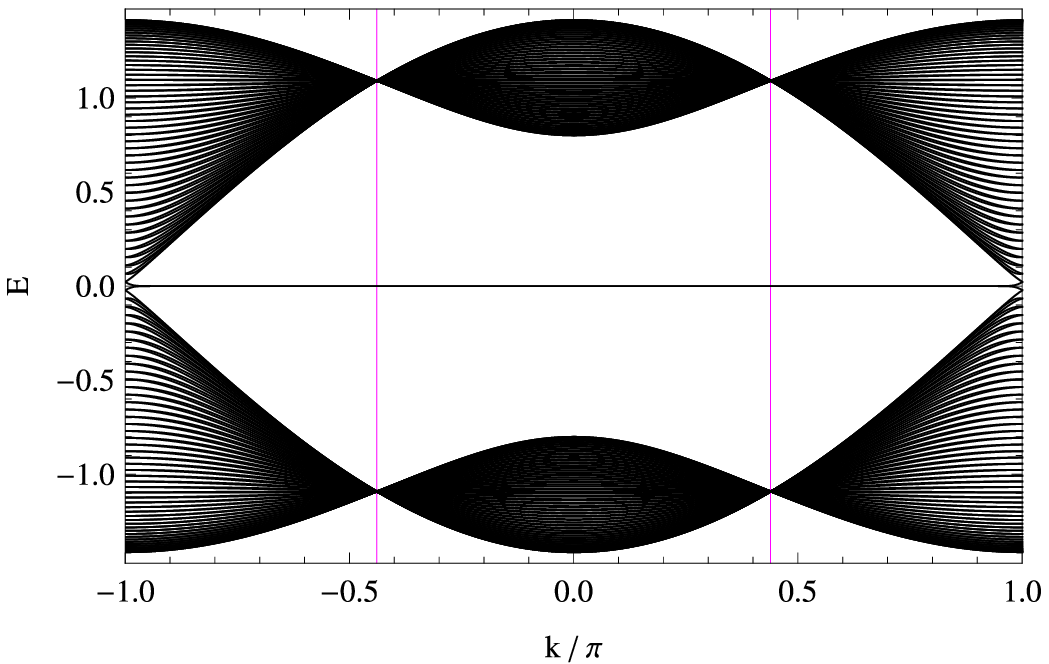}
\caption{$\lambda_-$ plotted as a function of $k$ at $A=1$
but for different values of $B$, i.e., 
$B=1, 0.6,0.4, 0.35, 0.2$ and $B=0$,
corresponding, respectively, the colors: red, green, blue, cyan, magenta
and orange (upper panel).
Edge spectrum at $\Delta=4B$ and $B=0.2$ ($A=1$, $N_r =100$, lower panel).
Two reference lines are at $k=\pm 0.438977...$,
the value of $k$ at which $\lambda_-$ vanishes at $B=0.2$.
The axes are suppressed in the lower panel so as to
highlight the completely flat edge modes.}
\label{lambda}
\end{figure}

In FIG. \ref{lambda}, the upper panel shows
$\lambda_-$ plotted as a function of $k$ at $A=1$
but for different values of $B$, 
naturally assuming $\Delta=4B$.
When $B$ is smaller than a critical value $B_c=0.35...$
$\lambda_-$ changes its sign (has a zero) at intermediate $k$.
\footnote{Clearly, at this value of $k$ the edge wave function is extremely localized,
i.e., onto a single row: $j=1$ or $j=N_r$.}
This continues to be the case even in the limit $B$ vanishes.
The lower panel shows
the edge spectrum at $\Delta=4B$, $B=0.2$ and $A=1$.
Note that 
the zero of $\lambda_-$ corresponds to the value of $k$ 
at which the bulk spectrum focuses onto a single point.

\subsection{Wave functions in special cases of parameters}

As we will describe in detail in the next subsection, our "recipe" for constructing 
the {\it exact} edge wave function, and simultaneously its spectrum,
lies in "extrapolating" the exact solution available at $\Delta=4B$
to a general value of $\Delta/B \in [0,8]$
(recall also FIG. \ref{zig_concept}).
To complete this program, we need to refer to some results of the numerical experiments
performed for a system of finite number of rows.
We have already seen the spectrum of such systems in FIG. \ref{spec_zig};
here we focus on the behavior of wave function, i.e.,
the behavior of $\psi_j$ as a function of $j$.

In the zigzag edge geometry, it is remarkable that (not only) the edge wave function
(but also the bulk wave function!)
has the following particular form:
\begin{equation}
\Psi=
\left[\begin{array}{c}
c_1 (1-i) \\ 
c_2 \\ 
c_3 (1-i) \\ 
c_4 \\
c_5 (1-i) \\
c_6 \\
\vdots
\end{array}
\right].
\label{evec_gen}
\end{equation}
when that eigenstate represents an edge mode, Eq. (\ref{evec})
further simplifies:
\begin{equation}
\Psi_\pm=\left[ 
\begin{array}{c}
c_2
\left[ \begin{array}{c} \alpha (1-i) \\ 1 \end{array} \right]
\vspace{0.2 cm} \\
c_4
\left[ \begin{array}{c} \alpha (1-i) \\ 1 \end{array} \right]
\vspace{0.2 cm} \\
c_6
\left[ \begin{array}{c} \alpha (1-i) \\ 1 \end{array} \right]
\vspace{0.2 cm} \\
\vdots \\
\end{array}
\right],
\label{evec_edge}
\end{equation}
i.e., for a given set of parameters $A$, $B$ (and $\Delta$) as well as for
a fixed $k$, $\alpha_j=c_{2j-1}/c_{2j}$ is a constant ($=\alpha$).
The ratio, on the other hand,
\begin{equation}
\rho_j = {c_{2j+2} \over c_{2j}},
\end{equation}
is a measure of, to what extent the edge mode is localized in the vicinity of
a boundary, say, at $j=1$.

\begin{figure}[htdp]
\includegraphics[width=8 cm]{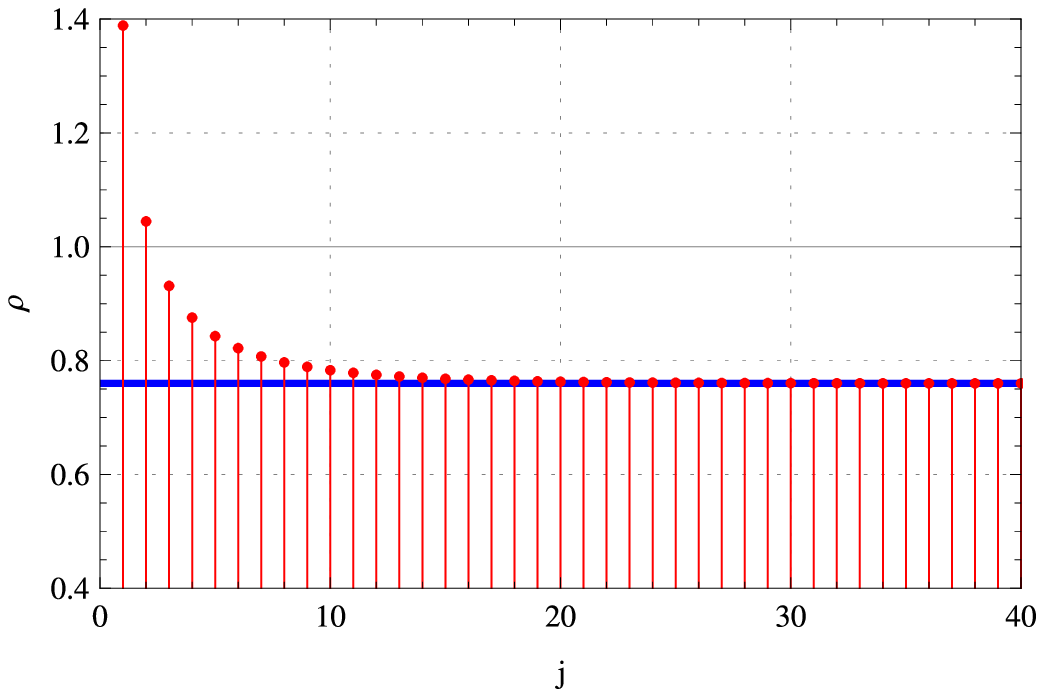}
\includegraphics[width=8 cm]{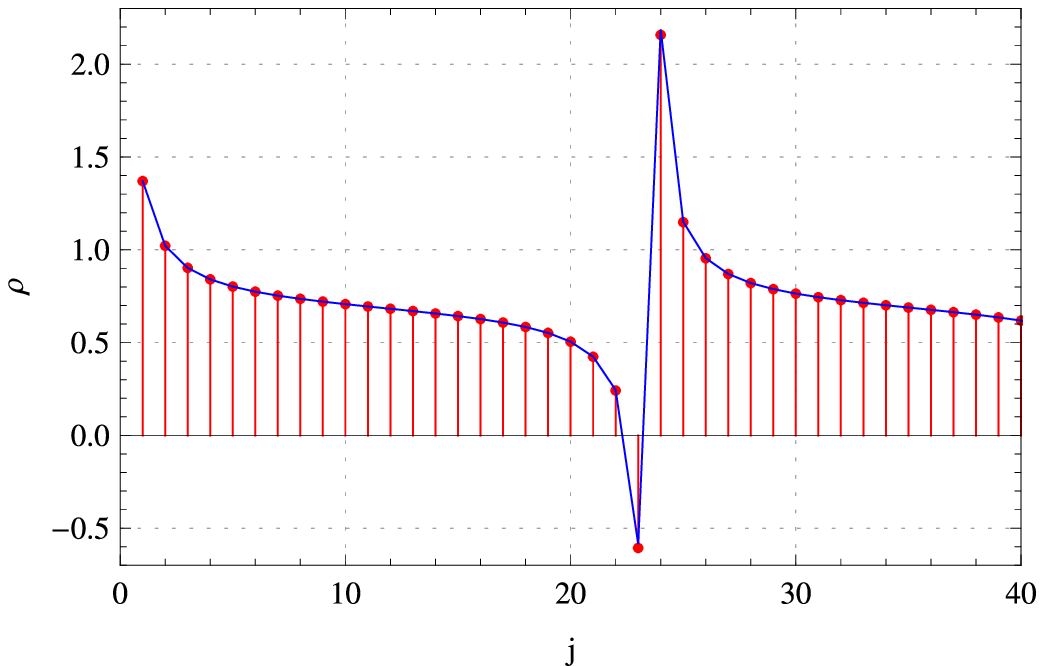}
\caption{The ratio,
$\rho_j = c_{2j+2} / c_{2j}$ ($j=1, 2, 3, \cdots$)
plotted (red points with filling to the horizontal axis)
at $k=0.05$
for $\Delta/B=0.25$ (upper panel) and $\Delta/B=0.30$ (lower panel).
$A=B=1$, $N_r=100$. 
In the upper panel, the blue line corresponds to the
"theoretical" value, $\rho_j=0.759908...$,
whereas
in lower panel, the plots are fitted by
a curve, $\rho_j=r \sin (j+1)\theta / \sin j\theta$, with the choice of parameters,
$r=0.691189...$, $\theta= 0.133449...$}
\label{rho_j}
\end{figure}

\begin{figure}[htdp]
\includegraphics[width=8 cm]{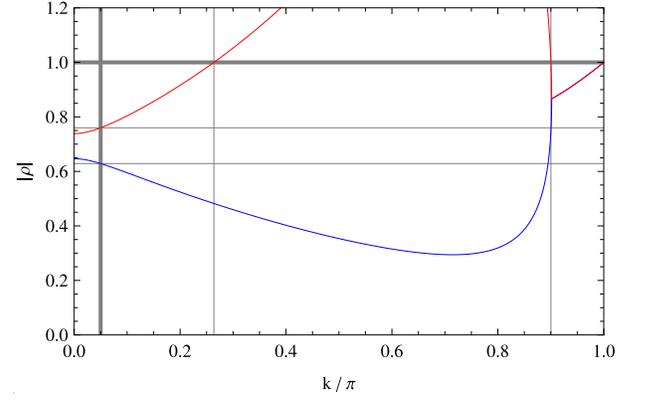}
\caption{Theoretical value (derived later) of $\rho$ (its magnitude, $|\rho|$) 
plotted as a function $k$ for $\Delta/B =0.25$ ($A=B=1$).
At $k/\pi =0.05$
there are two possible solutions for $\rho$:
$\rho_{-1}=0.759908...$ and $\rho_{-2}=0.628683...$,
the larger value of which determines large-$j$ behavior of $\rho_j = c_{2j+2} / c_{2j}$.
Merger with bulk occurs when $\rho_{-1}=1$, i.e.,
at $k/\pi = 0.263808...$
Close to the zone boundary ($k/\pi > 0.900237...$), 
reentrance of edge solution occurs (see FIG. \ref{reent_k} for details).}
\label{rho_th_D025}
\end{figure}
\begin{figure}[htdp]
\includegraphics[width=8 cm]{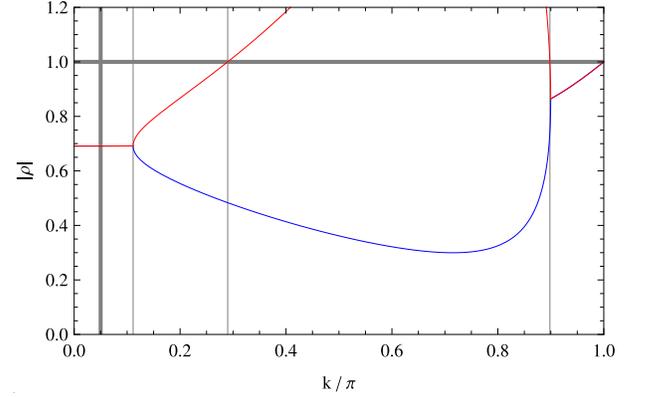}
\caption{Same as FIG. \ref{rho_th_D025} for $\Delta /B=0.3$ ($A=B=1$).
On the reference line at $k = 0.05$, 
$\rho \simeq 0.685038 \pm 0.0919993 i$ 
($|\rho|=0.691189...$, ${\rm Arg} [\rho] = \pm 0.133499...$).}
\label{rho_th_D030}
\end{figure}

We have extensively studied such characteristic behaviors of the edge wave function
in numerical experiments.
In FIG. \ref{rho_j},
results of such analyses are shown.
for the choice of parameters such that
$k/\pi=0.05$, $A=B=1$ and two different values of $\Delta/B$:
$\Delta=0.25$ and $\Delta=0.30$ for comparison.
At this value of $k/\pi=0.05$,
we first verified that the wave function $\psi_j$ takes indeed the form of
Eq. (\ref{evec_edge}),
with $\alpha_\pm$ approximately given by,
\begin{equation}
\alpha_+ = 0.687705...,
\ \ \ 
\alpha_- = - 0.727056...
\label{a_pm_k005}
\end{equation}
As in the straight edge case, for $k$ corresponding to an edge mode,
i.e., for such a state that are localized in the vicinity of either of the two boundaries,
this value of $\alpha_\pm$ is common practically to all $j$ in the strip,
as far as a finite amplitude exists.
For bulk states which are well extended into the interior of the sample,
the wave function takes no longer the form of Eq. (\ref{evec_edge}),
but keeps still a characteristic form as Eq. (\ref{evec_gen}).

In the two panels of FIG. \ref{rho_j},
$\rho_j$ is plotted as a function of $j$ for two different values of $\Delta/B$.
At $\Delta=0.25$, $\rho_j$ saturates at rows 
away enough from the boundary at $j=1$
(but not too close to the other edge, either),
to a value close to $\rho_j=0.759908...$ (blue line),
a value which is later "derived"
(see FIG. \ref{rho_th_D025}).
Let us assume,
\footnote{We leave formal derivation of Eq. (\ref{sol_zig_gen}) 
to a future publication.
Instead, we take it here as a plausible hypothesis fully justified by numerical experiments.}
as in the straight edge case,
that the wave function in the zigzag edge geometry
takes the following form:
\begin{equation}
\psi_j = \left[ c_{+1} \rho_1^j + c_{+2} \rho_2^j \right] u_+
\nonumber \\
+ \left[ c_{-1} \rho_1^{-j} + c_{-2} \rho_2^{-j} \right] u_-,
\label{sol_zig_gen}
\end{equation}
where the eigenspinors $u_\pm$ are always given by Eq. (\ref{u+-}),
the latter found analytically in the limit $\Delta/B \rightarrow 4$.
We have extensively verified the validity of this hypothesis
in numerical experiments.
The coefficients $c_{\pm 1,2}$ are susceptible of system's geometry.
Here, in a strip geometry, they satisfy,
\begin{equation}
c_{+1} = - c_{+2} \neq 0,\ \ \
c_{-1} = c_{-2} =0.
\label{c's_1}
\end{equation}
for one edge mode, and
\begin{equation}
c_{+1} = c_{+2} = 0,\ \ \
c_{-1} = - c_{-2} \neq 0
\label{c's_2}
\end{equation}
for the other.
Under this hypothesis,
such a behavior as seen in the upper panel of FIG. \ref{rho_j}
($\Delta=0.25$)
is interpreted as a consequence of two "real solutions" for $\rho$,
which are also {\it both} smaller than unity
(cf. FIG. \ref{rho_th_D025}).

On the other hand,
at $\Delta=0.30$ (in the lower panel of  FIG. \ref{rho_j})
$\rho_j$ shows an oscillatory behavior.
This implies, with the same hypothesis as above,
i.e., the wave function,
$\psi_j$, given as in Eq. (\ref{sol_zig_gen}),
with the choice of coefficients as Eqs. (\ref{c's_2}),
a pair of complex solutions for $\rho$.
Indeed, the plots at $\Delta=0.30$ are nicely fitted by a curve, of the form, 
\begin{equation}
\rho_j=|\rho| {\sin (j+1)\theta \over \sin j\theta} = r \left[ \cos \theta + \sin \theta  \cot (J\theta) \right],
\end{equation}
with the choice of parameters,
$|\rho|=0.691189...$ and $\theta= 0.133499...$,
which will be also (a posteriori) justified
(see FIG. \ref{rho_th_D030}).

\begin{figure}[htdp]
\includegraphics[width=8 cm]{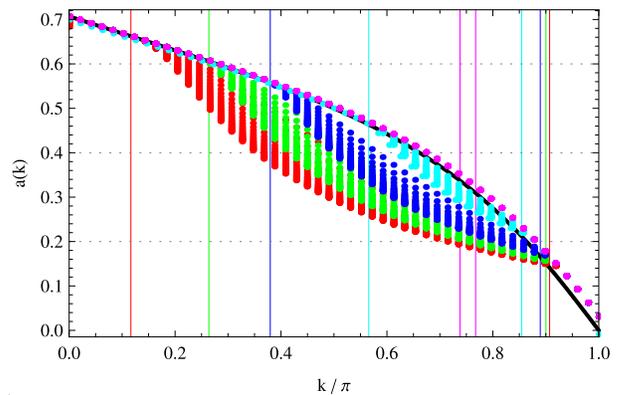}
\caption{$\alpha_j=c_{2j-1}/c_{2j}$ ($j=1,2,3,\cdots$) calculated in a strip geometry (here, $N_r=100$)
is plotted against the theoretical curve for $\alpha_-(k)$ 
(shown in black)
given in Eq. (\ref{a+-})
up to $j=N_r/4$ for the lowest-energy
eigenmode at different values of 
$\Delta/B=0.1, 0.25, 0.5, 1, 1.35$ ($A=B=1$ fixed), 
corresponding, respectively, to the colors:
red, green, blue, cyan and magenta.
Reference lines indicate the regime of $k$ in which the edge modes disappear
at the given value of $\Delta/B$.}
\label{ak}
\end{figure}

Comparing these two contrasting cases, notice that
the coefficients $\alpha_\pm$, 
estimated to be such as Eq. (\ref{a_pm_k005}),
are common to the two cases, i.e., independent of $\Delta/B$.
One can indeed verify (by changing the parameter $\Delta/B$ in numerical experiments)
that the eigenspinors $u_\pm$ remains always the same,
as far as the state describes an edge mode (see FIG. \ref{ak});
only $\rho_j$ changes as a function of $\Delta/B$.

In FIG. \ref{ak}, $\alpha_j=c_{2j-1}/c_{2j}$ ($j=1,2,3, \cdots$) is plotted for the lowest-energy
eigenmode $\Psi_0$ (in the upper band) at different values of $\Delta/B$.
One can see that in the range of $k$ at which
$\Psi_0$ is expected to represent an edge mode
the plotted points fall roughly on the theoretical curve for $\alpha_-(k)$ --- cf. Eq. (\ref{a+-}) ---
apart from a small disagreement close to the zone boundary ($k=\pi$).
This is indeed a key discovery allowing us to proceed to the next step, 
of extrapolating the earlier exact solution at $\Delta=4B$ to an arbitrary value
of $\Delta/B$.

\begin{figure*}
\begin{minipage}{8 cm}
\includegraphics[width=8 cm]{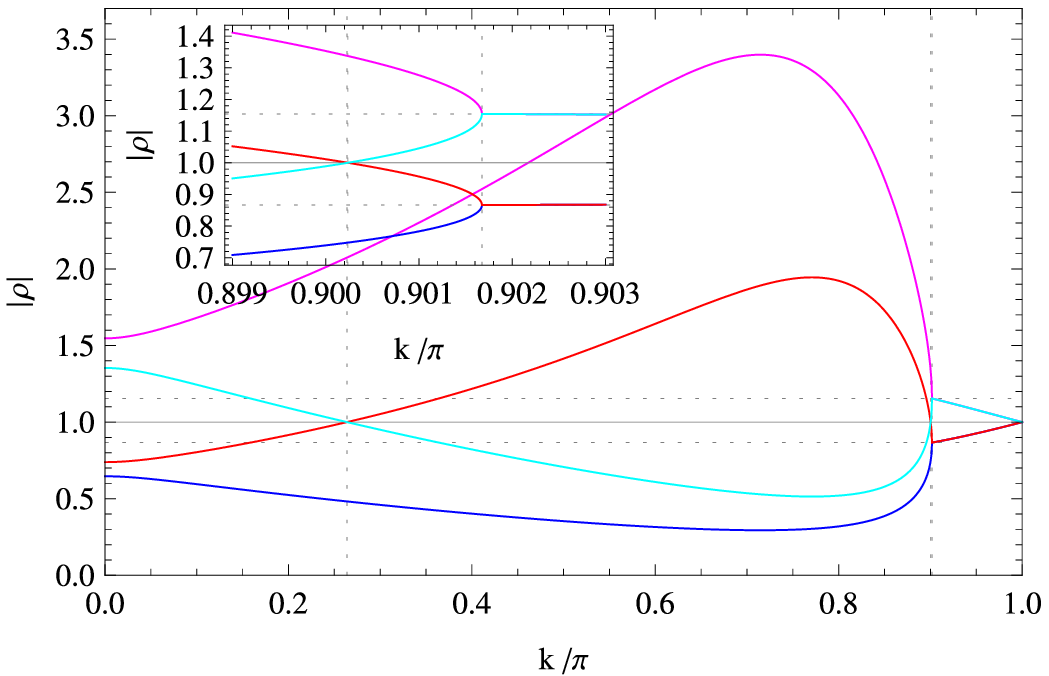}
\end{minipage}
\begin{minipage}{8 cm}
\includegraphics[width=8 cm]{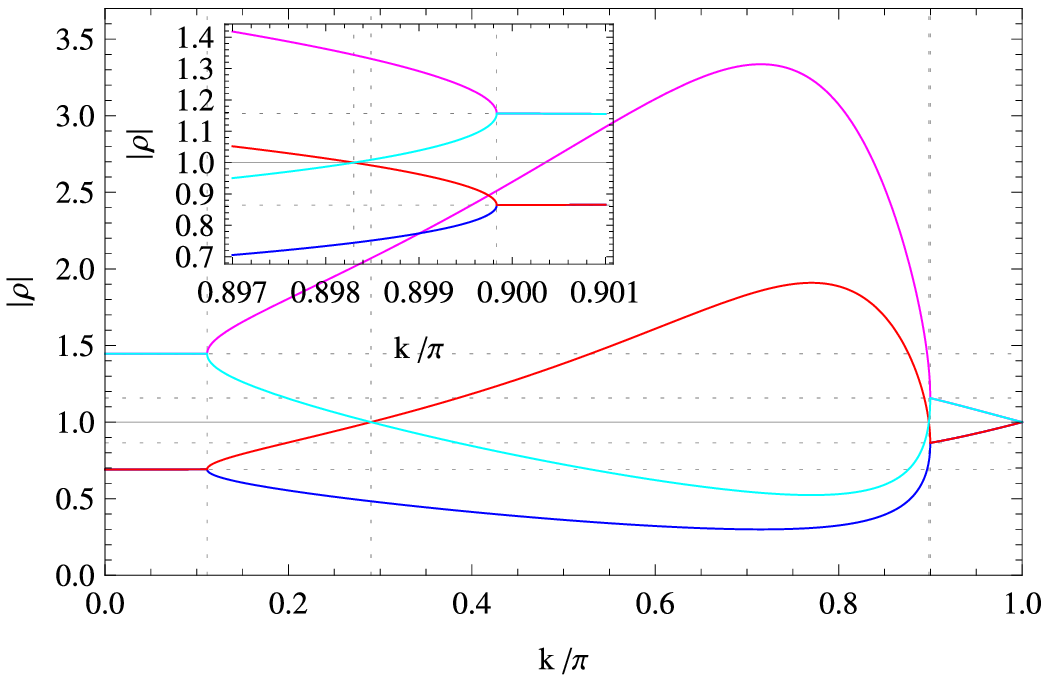}
\end{minipage}
\begin{minipage}{8 cm}
\includegraphics[width=8 cm]{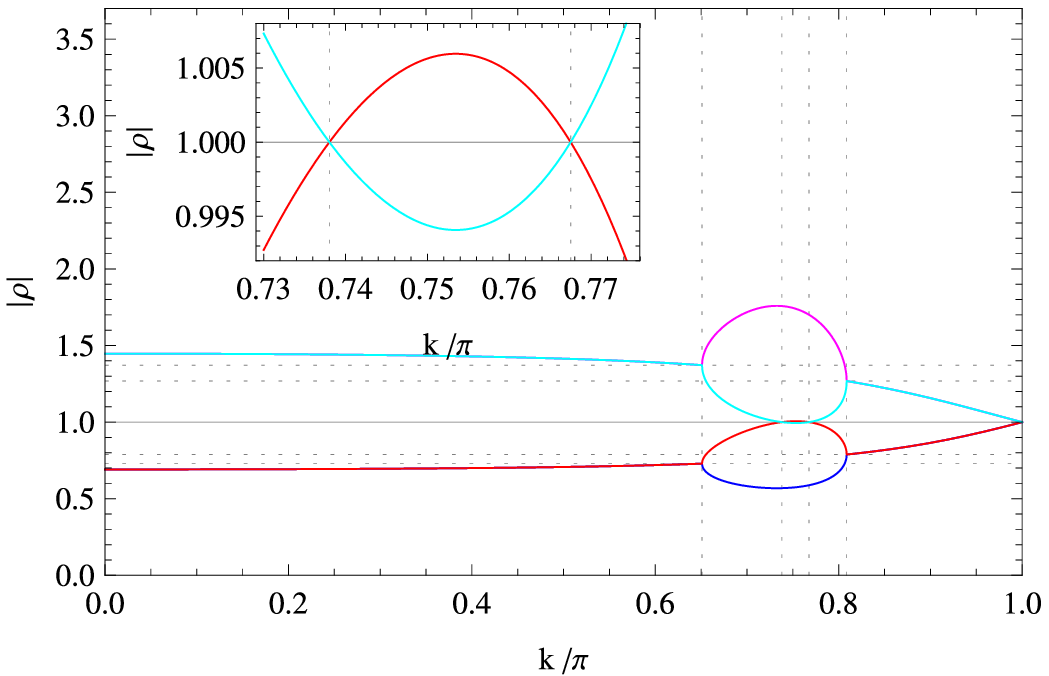}
\end{minipage}
\begin{minipage}{8 cm}
\includegraphics[width=8 cm]{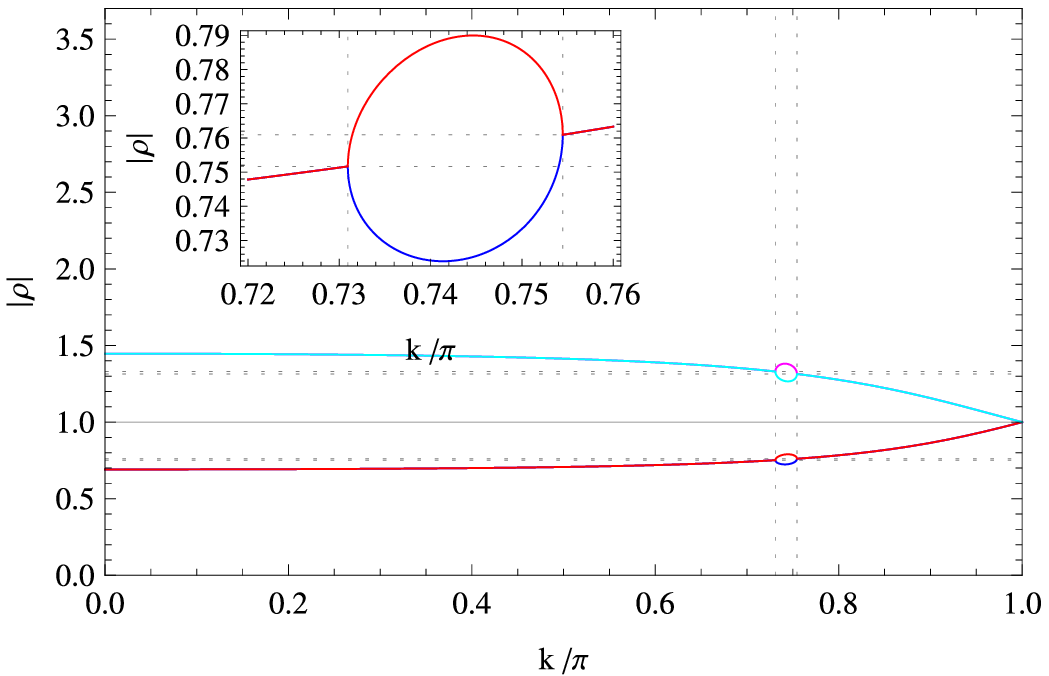}
\end{minipage}
\caption{Four solutions for $\rho$ 
(each curve corresponds to one solution).
Only the magnitude of such solution, which is generally a complex number,
i.e., $|\rho|$ is plotted as a function $k$ at different values of 
$\Delta /B=0.25, 0.3, 1.35$ and 1.45
($A=B=1$).}
\label{4sols}
\end{figure*}

\subsection{Derivation of exact edge wave functions}

Let us reformulate the recipe for constructing the {\it exact} edge wave function
and simultaneously its spectrum in the zigzag edge geometry, which has 
already been briefly outlined in the introduction (recall also FIG. \ref{zig_concept}).

\begin{enumerate}
\item 
We have seen in the previous subsection that
the edge wave function $\Psi$ in the zigzag edge geometry always takes,
as far as it describes a {\it localized edge mode},
the form of Eq. \ref{evec_edge} with a parameter $\alpha_\pm$
depending only on $k$ (and $A$, $B$).
All our numerical data agree with the hypothesis that $\alpha_\pm$,
consequently the reduced two-component eigenvector $u_\pm$,
is independent of $\Delta$.
\item
On the other hand, we know that the problem can be solved exactly at $\Delta =4B$.
We have seen, in particular, that the wave function $\Psi$ can be constructed
from the same set of spinors $u_\pm$ with a choice of parameters $\alpha_\pm$ 
given {\it analytically} as a function of $k$ in Eq. (\ref{a+-}).
\end{enumerate}

Taking also into account the fact that
the edge modes, gapless at $k=0$ and characterizing the topological insulator,
evolves continuously to the completely flat edge mode at $\Delta =4B$,
one can deduce, from these two observations, that the solution of the
eigenvalue equation for $\psi_0$, i.e., Eq. (\ref{reduced}) for an {\it arbitrary} $\Delta$
should be given, indeed, by $u_\pm$, defined as in Eq. (\ref{u+-}),
with the parameter $\alpha_\pm (k)$ obtained analytically in the limit:
$\Delta = 4B$ (recall FIG. \ref{ak}).

Thus, for a general value of $\Delta$, only $\rho$ and $E$ are unknown
(recall that the edge spectrum is no longer flat for a general $\Delta$).
But, clearly, they are solutions of
\begin{equation}
\left[
(\Delta -4B)\sigma_z + \rho\Gamma + {1\over\rho}\Gamma^\dagger 
\right]
\left[
\begin{array}{c}
\alpha (1- i)\\
1
\end{array}
\right]
= E 
\left[
\begin{array}{c}
\alpha (1- i)\\
1
\end{array}
\right],
\label{mao-chan}
\end{equation}
where the $2\times 2$ matrix $\Gamma$ is given explicitly as
Eq. (\ref{gamma_zig_mat}).
\footnote{Inspecting the explicit form of Eq. (\ref{mao-chan}) and (\ref{gamma_zig_mat}), 
notice that $1\pm i$ factors out. 
So all the coefficients become real.}
To find {\it exactly} the value of $\rho$ and $E$, one has only to solve 
this set of equations, and at the end of the calculation, substitute the
analytic expression for $a$, i.e., $\alpha_\pm$ given in Eq. (\ref{a+-})
obtained in the limit of $\Delta = 4B$.
Clearly, Eqs. (\ref{mao-chan}) are a set of coupled equations,
linear in $E$ and quadratic in $\rho$.
We expect, therefore, two sets of solutions for $(\rho, E)$,
which are given as a function of $a$.
To each of these two sets of solutions, one substitutes either $\alpha=\alpha_+$
or $\alpha=\alpha_-$.
There exist, therefore, four sets of solutions, in general.

Unfortunately, the analytic formula for these four sets of solutions are too lengthy
to be shown here.
Instead, we plotted these four solutions in FIG. \ref{4sols}, for different values of 
$\Delta /B$.

\begin{figure}[htdp]
\includegraphics[width=8 cm]{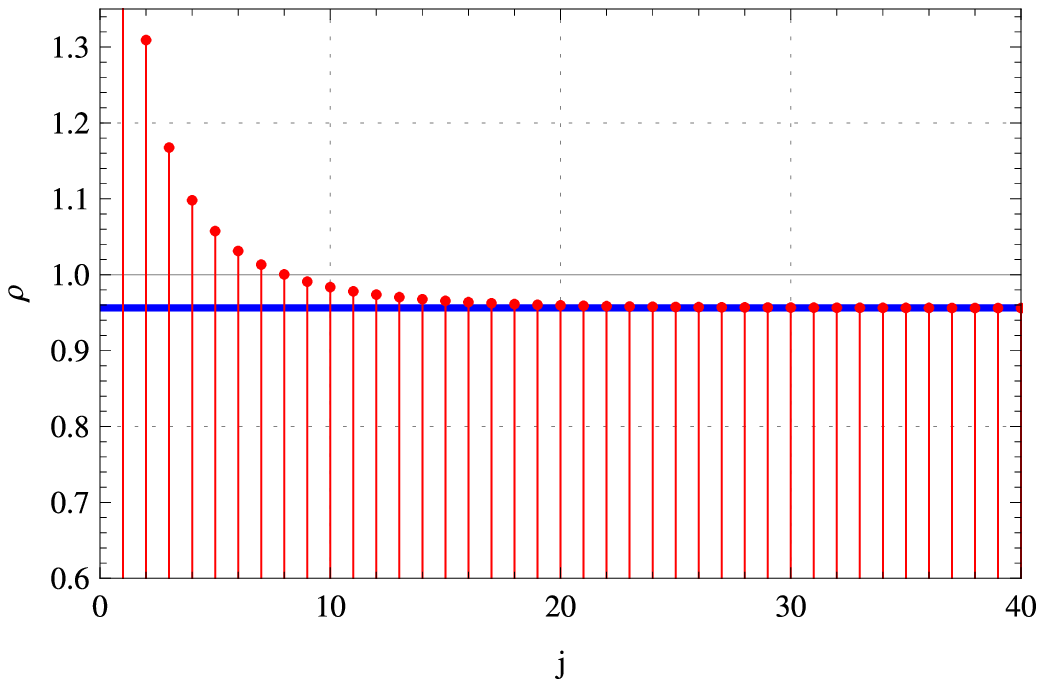}
\includegraphics[width=8 cm]{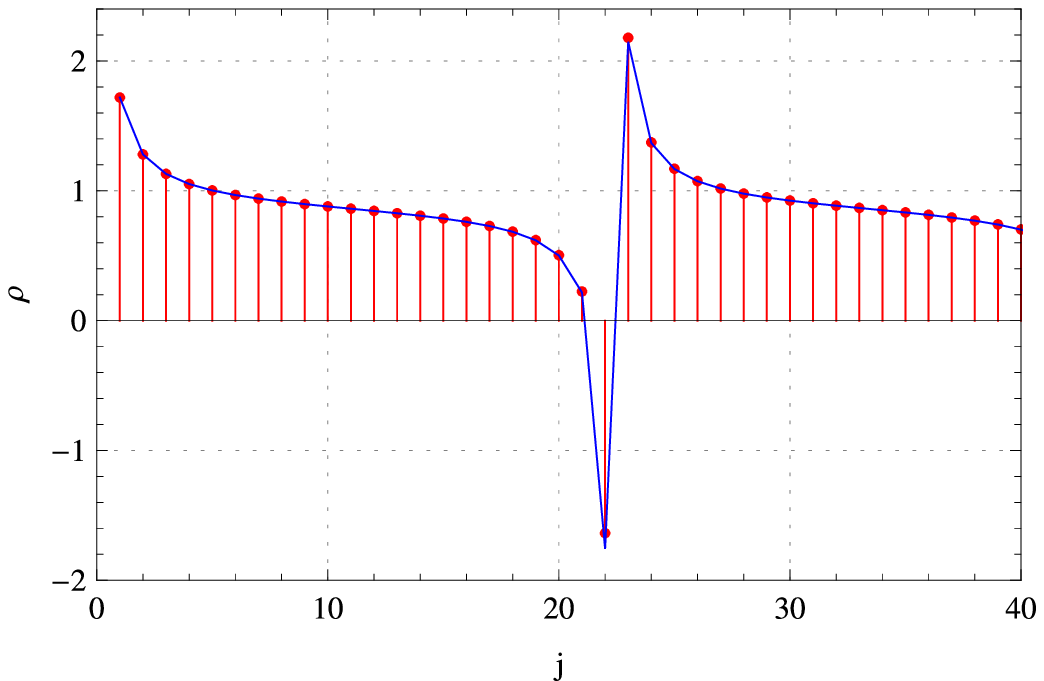}
\caption{Reentrant edge modes I:
$\rho_j$ plotted as a function of $j=1,2,3,\cdots$ at $A=B=1$ and $\Delta=0.25$
(as in FIG. \ref{rho_j}, upper panel), but for $k/\pi =0.901$ (upper panel) and 
$k/\pi =0.903$ (lower panel).
In the upper panel, the blue line corresponds to the theoretical value,
$\rho=0.956432$
whereas, in the lower panel
the plots (in red) are fitted by the curve, 
$\rho_j=r \sin (j+1)\theta / \sin j\theta$, with the choice of parameters,
$r=0.867804$, $\theta= 0.140687$.}
\label{reent_j}
\end{figure}

\begin{figure}[htdp]
\includegraphics[width=8 cm]{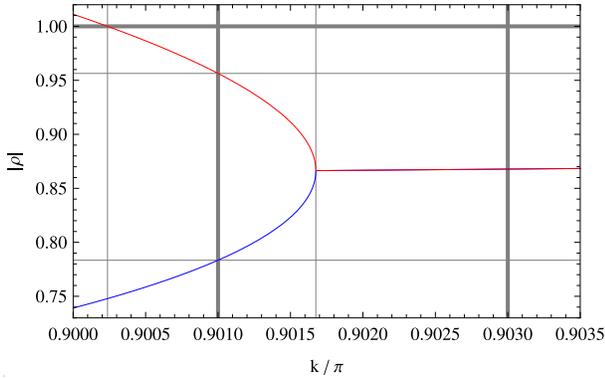}
\caption{(Theoretical value of) $|\rho|$ plotted as a function $k$ at $\Delta =0.25$
(again, as in FIG. \ref{rho_j}, upper panel).
Enlarged picture for $k$ close to the zone boundary: $k/\pi \sim 0.9$
The reentrance of edge solution occurs at $k/\pi = 0.900237...$, whereas
two real solutions for $\rho$ is possible when $k/\pi < 0.901675...$
Two reference lines are at $k/\pi =0.901$ and at $k/\pi=0.903$,
on which $\rho_j$ was plotted in FIG. \ref{reent_j}.
The value of $\rho$ on these lines are,
$\rho=0.783429...$ and $\rho=0.956432...$ on $k/\pi = 0.901$
(case of real solutions),
whereas
$\rho\simeq 0.859242 \pm 0.121601 i$ on $k/\pi = 0.903$.}
\label{reent_k}
\end{figure}

\begin{figure}[htdp]
\includegraphics[width=8 cm]{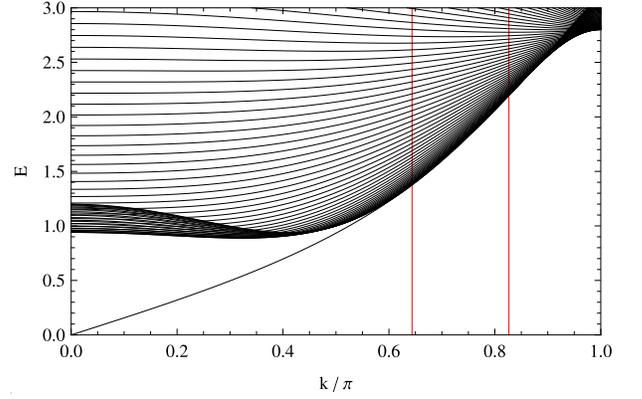}
\caption{Merger of the edge mode with bulk continuum and
"absence" of reentrance in the spectrum.
As for the latter, it turns out later that
binding energy of the edge mode is too small to be seen at this scale
(see FIG. \ref{en_bind}).
This is an enlarged image of the spectrum in the zigzag edge geometry
as shown in FIG. \ref{spec_zig}.
Here, the parameters are chosen such that
$\Delta /B=1.2$, $A=B=1$, $N_r =100$.
Two reference lines at $k/\pi = 0.643658... \equiv k_{c1}/\pi$ and
at $k/\pi = 0.826568... \equiv k_{c2}/\pi$ introduce
three different momentum regions:
i) $k/\pi \in [0, k_{c1}/\pi]$, ii) $k/\pi \in [k_{c1}/\pi, k_{c2}/\pi]$ and
iii) $k/\pi \in [k_{c2}/\pi,1]$,
corresponding, respectively, to i) the ordinary edge,
ii) the bulk and iii) the reentrant regimes.}
\label{spec_zig_enl}
\end{figure}

\subsection{Reentrant edge modes}

{\it Reentrance} of the edge mode is another characteristic feature 
of the edge mode of zigzag geometry,
and occurs close to the zone boundary, $k/\pi =1$, 
when $\Delta/B$ is not too large: $\Delta/B < 1.354...$.
Very remarkably,
the spectrum looks completely "innocent" when this occurs, 
i.e., the edge mode, say, 
the lowest energy ($=E_0$) mode in the upper band looks
almost completely degenerate with the bottom of the (bulk) spectrum ($=E_1$),
in this regime of $k$
(see FIG. \ref{spec_zig_enl}).
Existence of an edge mode of such specific character 
is, on the other hand, nothing exceptional in the zigzag edge geometry.
At a value of $\Delta$, e.g., $\Delta /B = 0.25$ or $\Delta /B = 0.30$
as in FIG. \ref{rho_j},
such reentrant edge modes are indeed existent.
If one focuses on the wave function of, say, 
the lowest energy mode in the upper band,
after touching the lower band at $k=0$, 
it continues to be spatially localized when $k$ is small enough, 
but as the spectrum merges with the bulk continuum, the wave function
also starts to penetrate into the bulk.
However, close to the zone boundary, it starts to be localized again.
This is what we call the reentrance of edge modes.

Figs. \ref{reent_j} and \ref{reent_k} 
highlight the behavior of such reentrant edge modes,
naturally in a different regime of $k$ from, say, FIG. \ref{rho_j}.
Fig. \ref{reent_j} shows the behavior of $\rho_j$
at $k/\pi=0.901$ and $k/\pi=0.903$
for $\Delta =0.25$.
At this value of $k$,
the wave function $\psi_j$ takes always the form of Eq. (\ref{evec_edge}),
but with a different set of parameters for $\alpha_\pm$
from the case of FIG. \ref{rho_j}, upper panel, 
since it still depends on $k$.
The two plots for $\rho_j$ in Fig. \ref{reent_j} show two typical behaviors
of the edge wave function, i.e.,
one corresponding to real and the other to complex solutions for $\rho$.
As we have extensively studied in the case of ordinary edge modes
(appearing at $k/\pi \ll 1$),
the two contrasting behaviors of $\rho_j$ as a function $j$ (in FIG. \ref{rho_j})
are naturally understood by referring to the theoretical curve of $\rho$ 
as a function of $k$, e.g., such as the one shown in FIG. \ref{rho_th_D025}.

What is rather remarkable here, in the case of Fig. \ref{reent_j}, is that
this  crossover between real and complex solutions occurs within 
a tiny change of $k$, i.e.,
from $k/\pi=0.901$ in the upper panel to $k/\pi=0.903$ in the lower panel.
This drastic change is, however, quite reasonable from the viewpoint of
FIG. \ref{reent_k}.
The upper panel of Fig. \ref{reent_j} shows a monotonic decay, 
which converges asymptotically to a single  exponential decay.
This is consistent with the behavior of theoretical curve for $\rho$ 
as a function of $k$ in FIG. \ref{reent_k}.
The latter implies two real solutions for $\rho$ at $k/\pi=0.901$:
$\rho=0.783429...$ and $\rho=0.956432...$
The latter coincides with the value of $\rho_j$ in FIG. \ref{reent_j}
at which it saturates.
At $k/\pi=0.903$, on the other hand, 
the plots for $\rho_j$ are nicely fitted by the curve, 
$\rho_j=r \sin (j+1)\theta / \sin j\theta$, with the choice of parameters,
$r=0.867804...$ and $\theta= 0.140687...$
(see the lower panel of Fig. \ref{reent_j}).
This is a clear fingerprint that 
the reentrant edge mode at this value of $k$
corresponds to a pair of complex solutions for $\rho$.

\begin{figure}[htdp]
\includegraphics[width=8 cm]{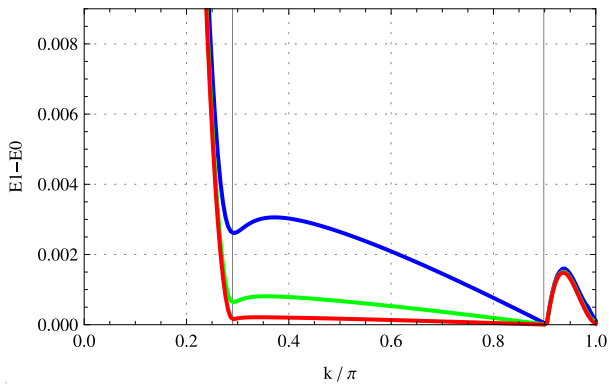}
\caption{Binding energy of the reentrant edge mode:
$E_1-E_0$ is plotted as a function of $k$ at
$\Delta = 0.3 B$, $A=B=1$.
Different curves correspond to different size (width) of the system:
$N_r=100$ (blue), 200 (green) and 300 (red).
The two reference lines are placed at $k/\pi=0.289936\equiv k_{c1}/\pi$ 
and $k/\pi=0.8983 \equiv k_{c2}/\pi$ 
($k \in [k_{c1}, k_{c2}]$ corresponds to the bulk regime).
The plots reveal an extremely small but still a finite binding energy of 
the reentrant edge modes.}
\label{en_bind}
\end{figure}

Does the reentrant edge mode really have zero binding energy?
In order to address this question,
we (re)plotted the energy spectrum
($E_1-E_0$, to be precise) but in an enlarged scale roughly by one thousand times
in FIG. \ref{en_bind}.
First, for an "ordinary" edge state, occurring at 
$0 < |k|/\pi < 0.289936 \equiv k_{c1}/\pi$ 
the value $E_1-E_0$ is much above the threshold at this scale.
$E_1-E_0$ takes a value of order $\sim 1$ for such ordinary edge state.
As for the reentrant edge mode,
FIG. \ref{en_bind} reveals that it has indeed an extremely small but still 
a finite binding energy.
Notice different behaviors of $E_1-E_0$ as a function of $N_r$ 
in the bulk and reentrant regions of $k$.
The former (the latter) corresponds to
$k/\pi \in [k_{c1}/\pi, k_{c2}/\pi \equiv 0.8983...]$ 
($k/\pi \in [k_{c2}/\pi,1]$).
In the bulk region $E_1-E_0$ is expected to vanish in the thermodynamic limit.
FIG. \ref{en_bind} shows indeed that the binding energy of 
reentrant edge mode, $E_1-E_0 \sim 0.001$,
is thousand times smaller 
than that of the ordinary edge state.
This implies the appearance of an extremely small energy scale
which was not existing in the original Hamiltonian
(cf. Kondo effect).

The reentrance of edge mode is indeed
a unique feature, in its contrasting properties in real and momentum space,
and in the appearance of an extremely small energy scale.

\section{Conclusions}

We have highlighted in this paper various unique properties of helical edge modes 
in $Z_2$ topological insulator.
We have extensively investigated a lattice version of the BHZ model,
under different edge geometries.
One of the specific characters of BHZ model is that
the spin Hall conductance in the bulk
changes its sign in the middle of topological phase
(at $\Delta = 4 B$),
i.e., $\sigma^{(s)}_{xy}=\pm e/(2 \pi)$,
respectively, for $0<\Delta /B<4$ and for $4<\Delta /B<8$,
though both represent a non-trivial value.
From the viewpoint of bulk-edge correspondence,
this information should be also encoded in the edge theory.
We have seen that
the change of $\sigma^{(s)}_{xy}$ manifests
in a very different way in the $(1,0)$- (straight) and $(1,1)$- (zigzag) edge geometries.
In the $(1,0)$-edge case,
the edge spectrum changes its global structure
in the two parameter regimes, i.e.,
the main location of the mode moves from the zone center for $\Delta /B < 4$, 
to the zone boundary for $\Delta /B > 4$.
As a result, the group velocity at the intersection with Fermi level reverses its sign,
leading to change of the sign in Landauer conductance at the edge.
In the $(1,1)$-edge case, on the other hand,
the edge spectrum is symmetric w.r.t. $\Delta = 4 B$,
i.e., neither change of the position of gap closing, nor the reversal of group velocity
at $\Delta = 4 B$.
The change of $\sigma^{(s)}_{xy}$ is here encoded in the swapping of 
left- and right- going edge modes of the same spin.

Much of our focuses has been on the analysis of
the zigzag or  $(1,1)$-edge geometry,
the latter showing,
as a consequence of specific way in which
the bulk topological structure is projected onto the 1D edge,
a number of unique features, such as
the completely flat edge spectrum at $\Delta = 4 B$,
and the reentrance of edge modes.
We have also shown, here in a half-empirical way, that 
the exact edge wave function for zigzag edge geometry
can be constructed,
by extrapolating the solution at $\Delta = 4 B$.
The reentrant edge mode, 
though sharing much of its characteristics with the usual edge mode in real space, 
introduces a new extremely small energy scale which was absent in the original BHZ model.

\acknowledgments
KI, AY and AH have been much benefited from useful discussions with Jun Goryo on the bulk/edge
correspondence.
KI also acknowledges Christoph Br\"une, Hartmut Buhmann and Laurence Molenkamp for 
their detailed explanation of the experimental situations in STCM (Kyoto), NGSS-14 (Sendai) and
QHSYST10 (Dresden) conferences.
KI and AY are supported by KAKENHI 
(KI: Grant-in-Aid for Young Scientists B-19740189,
AY: No. 08J56061 of MEXT, Japan).

\appendix

\section{Edge solution in the long-wave-length limit \cite{ShenNiu,Shen+,Sonin}}

Let us first recall that
the eigenvector $|\bm d (\bm k) \pm\rangle$,
which has appeared in Eq. (\ref{dk_pm}),
is a standard SU(2) spinor, here chosen to be single-valued.
An eigenvector, corresponding to a positive energy eigenvalue $E>0$,
is $|\bm d (\bm k) +\rangle$ with $\bm k$ satisfying $E=E(\bm k)$, 
and represented as, 
\begin{eqnarray}
|\bm d (\bm k) +\rangle &=&
\left[
\begin{array}{c}
e^{- i \phi} \cos (\theta/2) \\
\sin (\theta/2)
\end{array}
\right]
\\ \nonumber
&=&{1\over\sqrt{2d(d-d_z)}}
\left[
\begin{array}{c}
d_x - i d_y \\
d - d_z
\end{array}
\right].
\label{evec}
\end{eqnarray}
$\theta$, $\phi$ are polar coordinates in $\bm d$-space,
satisfying the relations such as,
\begin{equation}
\cos \theta = {d_z\over d},\ \
\cos \phi = {d_x\over \sqrt{d_x^2+d_y^2}}.
\end{equation}

In order to find an edge solution,
we focus on a solution of the form, \cite{ShenNiu}
\begin{equation}
\psi (k_x, y) = \phi_\kappa (k_x)\ e^{\kappa y},
\label{evec_kappa}
\end{equation}
say, in a semi-infinite plane: $y<0$.
The spatial dependence in the $y$-direction
can be taken into account by applying Pierls substitution:
$k_y \rightarrow -i\partial/\partial y$
to $k_y$'s in $h(\bm k)$.

The eigenenergy of such a solution is obtained
by a simple replacement:
$k_y\rightarrow -i\kappa$ in Eq. (\ref{wine}), i.e.,
\begin{equation}
E^2=\Delta^2+\left( A^2-2B\Delta \right) (k_x^2-\kappa^2) +B^2 (k_x^2-\kappa^2)^2.
\end{equation}
This can be regarded as a quadratic equation w.r.t. $\kappa^2$.
Its two solutions are,
\begin{eqnarray}
\kappa^2=k_x^2&+&{A^2-2B\Delta \over 2B^2}
\nonumber \\
&\pm&{1\over 2B^2}\sqrt{A^2\left(A^2-4B\Delta\right) +4B^2E^2}.
\label{sol_kappa}
\end{eqnarray}
For a given set of $k_x$ and $E$,
there are two possible values for $\kappa^2$, or equivalently,
four possible values for  $\kappa$.
Of course, in a semi-infinite plane, say, $y<0$
the edge solution of the form, Eq. (\ref{evec_kappa})
should decay as $y\rightarrow -\infty$, so only two of such solutions are
relevant.

We expect that the edge spectrum behaves as $E \rightarrow 0$ 
in the limit of$k_x \rightarrow 0$.
Let us parametrize the two solutions in this limit as
\begin{equation}
\kappa^2 =\left[\kappa_\pm^{(0)}\right]^2\equiv P\pm Q,
\label{kappa0}
\end{equation}
where
\begin{equation}
P={A^2-2B\Delta \over 2B^2},\ \ \
Q={1\over 2B^2}\sqrt{A^2\left(A^2-4B\Delta\right)}.
\label{ab}
\end{equation}
When $0<\Delta<A^2/(4B)\equiv \Delta_1$,
$P>0$ and $Q$ is real.
Since $|a|>b$ as far as $b$ is real,
Eq. (\ref{sol_kappa}) represents two positive solutions for $\kappa^2$,
i.e., the wave function represented by Eq. (\ref{evec_kappa}) 
shows simple exponential damping.
On contrary, we expect a pure imaginary solution for $\kappa$
for an extended state in the bulk.
On the other hand,
when $\Delta_1<\Delta<\Delta_0=A^2/(2B)$,
$P$ is always positive but $Q$ becomes purely imaginary.
Thus two solutions for $\kappa$ become complex numbers
conjugate to each other.
In this case,
the wave function represented by Eq. (\ref{evec_kappa})
shows damped oscillation.

The corresponding eigenvector is obtained by the same replacement
$k_y\rightarrow -i\kappa$, here in Eq. (\ref{evec}), i.e.,
\begin{equation}
\phi_\kappa (k_x)=
\left[ 
\begin{array}{c} 
u_\kappa\\ 
v_\kappa 
\end{array} 
\right]
=\left[
\begin{array}{c}
A (k_x - \kappa) \\
E-d_z (\kappa)
\end{array}
\right]
\sim
|\bm d (k_x, -i\kappa) +\rangle 
\end{equation}
where $d_z (\kappa)=\Delta+B(k_x^2-\kappa^2)$.
For a given value of $k_x$ and $E>0$,
we thus have identified two solutions characterized by different values
of $\kappa$.
In order to construct a general solution in the presence of a boundary,
we need to take a linear combination of these two solutions,
i.e.,
\begin{eqnarray}
\psi (y)&=&
c_+ \phi_{\kappa_+} e^{\kappa_+ y} + 
c_- \phi_{\kappa_-} e^{\kappa_- y}
\\ \nonumber
&\equiv& c_+
\left[ 
\begin{array}{c} 
u_+\\ 
v_+ 
\end{array} 
\right]
e^{\kappa_+ y} 
+  c_- 
\left[ 
\begin{array}{c} 
u_-\\ 
v_- 
\end{array} 
\right]
e^{\kappa_- y},
\end{eqnarray}
where $u_\pm$, $v_\pm$ are short-hand notations for
 $u_{\kappa_\pm}$ and $v_{\kappa_\pm}$.

We now fix the boundary condition at $y=0$,
which we choose to be,
\begin{equation}
\psi(y=0)=
\left[
\begin{array}{cc}
u_+ & u_- \\
v_+ & v_-
\end{array}
\right]
\left[
\begin{array}{c}
c_+\\
c_-
\end{array}
\right]=
\left[
\begin{array}{c}
0\\
0
\end{array}
\right],
\label{bc}
\end{equation}
which implies the following secular equation:
\begin{equation}
\det\left[
\begin{array}{cc}
A (k_x - \kappa_+) & A (k_x - \kappa_-) \\
E - d_z (\kappa_+) & E - d_z (\kappa_-)
\end{array}
\right]=0
\end{equation}
This leads to,
\begin{equation}
E (k_x)= \Delta - B \kappa_+ \kappa_-
+ B k_x(\kappa_+ +\kappa_-) - B k_x^2.
\label{req_bc2}
\end{equation}

We have thus identified the two basic equations,
Eqs. (\ref{sol_kappa}) and (\ref{req_bc2}),
for determining the energy spectrum
$E=E(k_x)$.

Let us check whether this solution contains the edge modes.
We expect that the edge spectrum bahaves,
as $k_x \rightarrow 0$, $E \rightarrow 0$.
Recall that in this limit, Eq. (\ref{sol_kappa})
reduces to Eqs. (\ref{kappa0}) and (\ref{ab}).
Eq. (\ref{req_bc}) is also simplified in this limit, as
\begin{equation}
E = E (0) = \Delta - B \kappa_+^{(0)} \kappa_-^{(0)}.
\label{req_bc0}
\end{equation}
Focusing on the case, $\kappa_\pm (0)>0$ and $B,\Delta>0$, 
and using the parameterization in Eqs. (\ref{kappa0}) and (\ref{ab}),
one can readily verify,
\begin{equation}
\kappa_+^{(0)} \kappa_-^{(0)}= \sqrt{P^2-Q^2}={\Delta\over B}.
\end{equation}
Thus Eq. (\ref{req_bc0}) is safely satisfied.

How about the first order corrections? 
i.e., contributions of order ${\cal O} (k_x)$ to
the energy spectrum, $E=E(k_x)$.
First note that there is no ${\cal O} (k_x)$-correction to $\kappa_\pm$.
One can, therefore, safely replace, at this order, $\kappa_\pm$'s in
Eq. (\ref{req_bc}) with their values at $k_x \rightarrow 0$, $E \rightarrow 0$,
i.e.,
\begin{equation}
E = E (0) = \Delta - B \kappa_+^{(0)} \kappa_-^{(0)} + B k_x 
\left[ \kappa_+^{(0)} + \kappa_-^{(0)} \right].
\label{req_bc1}
\end{equation}
We have already seen that the first two terms cancel, whereas
\begin{equation}
\left[ \kappa_+^{(0)} + \kappa_-^{(0)} \right]^2=
2a+2\sqrt{P^2-Q^2}={A^2\over B^2}.
\end{equation}
Thus, the edge spectrum in the continuum limit is determined to be,
\begin{equation}
E (k_x)=\pm A k_x + {\cal O} (k_x^2).
\label{edge_cont}
\end{equation}
Remarkably, the slope of the edge spectrum depends only on 
a single parameter, $A$.
An interesting question is to what extent this conclusion is general?
If one calculates the edge spectrum, using a tight-binding model,
generally the results depend on the way edges of the sample are introduced 
with respect to the lattice.
In the case of zigzag edge, in particular,
apparently the edge spectrum does not converge to Eq. (\ref{edge_cont})
even in the long-wave-length limit: $k_x \rightarrow 0$.



\begin{thebibliography}{99}

\bibitem{HgTe_JPSJ} 
M. K\"{o}nig, H. Buhmann, L.W. Molenkamp, T.L. Hughes, C.X
Liu, X.L Qi and S.C Zhang, J. Phys. Soc. Jpn \textbf{77}, 031007 (2008).

\bibitem{KM_Z2} 
C.L. Kane and E.J. Mele, 
Phys. Rev. Lett. {\bf 95}, 226801 (2005).

\bibitem{KM_QSH} 
C.L. Kane and E.J. Mele,
Phys. Rev. Lett. {\bf 95}, 146802  (2005). 

\bibitem{BHZ}
B. A. Bernevig, T. L. Hughes and S.-C. Zhang, Science 314, 1757 (2006).

\bibitem{Laurens}
M. K\"onig, S. Wiedmann, C. Br\"une, A. Roth, H. Buhmann, L.W. Molenkamp,
X.-L. Qi and S.-C. Zhang, Science {\bf 318}, 766 (2007).

\bibitem{Neto} 
A. H. Castro Neto, F. Guinea, N. M. R. Peres, K. S. Novoselov and A. K. Geim,
Rev. Mod. Phys. {\bf 81}, 109 (2009).

\bibitem{W_PhD} 
K. Wakabayashi, PhD Thesis, University of Tsukuba, 2000.

\bibitem{Wen}
X.-G. Wen, Int. J. Mod. Phys. B {\bf 6}, 1711 (1992).

\bibitem{HG}
Y. Hatsugai, Phys. Rev. Lett. {\bf 71}, 3697 (1993);
Phys. Rev. B {\bf  48}, 11851 (1993).

\bibitem{TKNN} 
D. J. Thouless, M. Kohmoto, M.P. Nightingale and M. den Nijs, 
Phys. Rev. Lett. {\bf 49}, 405 (1982);
M. Kohmoto, Annals of Physics {\bf 160}, 343 (1985).

\bibitem{Oshikawa}
M. Oshikawa, Phys. Rev. B {\bf 50}, 17357 (1994).

\bibitem{ShenNiu}
B. Zhou, H.-Z. Lu, R.-L. Chu, S.-Q. Shen and Q. Niu,
Phys Rev. Lett. {\bf 101}, 246807 (2008).

\bibitem{Jackiw}
S. Deser, R. Jackiw and S. Templeton, Phys. Rev. Lett. {\bf 48}, 975 (1982).

\bibitem{Niemi}
A.J. Niemi and G.W. Semenoff, Phys. Rev. Lett. {\bf 51}, 2077 (1983).

\bibitem{Redlich}
A.N. Redlich, Phys. Rev. Lett. {\bf 52}, 18 (1984).

\bibitem{Kenzo} 
K. Ishikawa and T. Matsuyama, Nucl. Phys. {\bf B 280}, 523 (1987).  

\bibitem{HKW}
Y. Hatsugai, M. Kohmoto and Y.S. Wu, 
Phys. Rev. B {\bf 54}, 4898 (1996).

\bibitem{HgTe_band}
E.G. Novik, A. Pfeuffer-Jeschke, T. Jungwirth, V. Latussek,
C.R. Becker, G. Landwehr, H. Buhmann and L. W. Molenkamp,
Phys. Rev. B {\bf 72}, 035321 (2005).

\bibitem{nogo}
H.B. Nielsen and M. Ninomiya,
Phys. Lett. 105 B, 219;
Nucl. Phys. B {\bf 185} (1981) 20; ibid.
{\bf 193}, 173 (1981).

\bibitem{FuKane}
L. Fu, C.L. Kane, Phys. Rev. B {\bf 76}, 045302 (2007).

\bibitem{SM}
S. Murakami, Prog. Theo. Phys. Supp. \textbf{176}, 
279 (2008).

\bibitem{W}
M. Fujita, K. Wakabayashi, K. Nakada, and K. Kusak- 
abe: J. Phys. Soc. Jpn. {\bf 65}, 1920 (1996).

\bibitem{Haldane}
F.D.M. Haldane, Phys. Rev. Lett. {\bf 61}, 2015 (1988).

\bibitem{Shen+}
H.-Z. Lu, W.-Y. Shan, W. Yao, Q. Niu, and S.-Q. Shen, Phys. Rev. B {\bf 81}, 115407 (2010);
W.-Y. Shan, H.-Z. Lu, and S.-Q. Shen, New J. Phys. {\bf 12} 043048 (2010).

\bibitem{Sonin}
E.B. Sonin, arXiv:1006.5218.

\end{thebibliography}
\end{document}